\documentclass[letterpaper,12pt]{article}
\usepackage[letterpaper,text={17cm,23.7cm},centering]{geometry}
\usepackage[utf8]{inputenc}
\usepackage[english]{babel}
\usepackage{amsmath}
\usepackage{amssymb}
\usepackage{latexsym}
\usepackage[table]{xcolor}
\usepackage{pdflscape}
\usepackage[dvips]{graphicx}
\usepackage{epsfig}
\usepackage{authblk}
\usepackage{setspace}
\usepackage[normalem]{ulem}

\usepackage[sort&compress]{natbib}

\usepackage[labelfont=bf]{subcaption}

\DeclareTextSymbol{\degre}{T1}{6}

\DeclareCaptionLabelFormat{nobracket}{\textbf{#2}}
\DeclareCaptionSubType{figure}

\title{ Antibody-mediated cross-linking of gut bacteria hinders the spread of antibiotic resistance}
\author[1,2]{Florence Bansept}
\author[1,2]{Loïc Marrec}
\author[1,3]{Anne-Florence Bitbol}
\author[1,3]{Claude Loverdo}
\affil[1]{Sorbonne Université, CNRS, Laboratoire Jean Perrin (UMR 8237), F-75005 Paris, France}
\affil[2]{These authors contributed equally.}
\affil[3]{Corresponding authors: anne-florence.bitbol@sorbonne-universite.fr; claude.loverdo@sorbonne-universite.fr}
\date{\today}

\begin{document}



\maketitle


\setcounter{page}{1}

\section*{Abstract}

The body is home to a diverse microbiota, mainly in the gut. Resistant bacteria are selected for by antibiotic treatments, and once resistance becomes widespread in a population of hosts, antibiotics become useless. Here, we develop a multiscale model of the interaction between antibiotic use and resistance spread in a host population, focusing on an important aspect of within-host immunity. Antibodies secreted in the gut enchain bacteria upon division, yielding clonal clusters of bacteria. We demonstrate that immunity-driven bacteria clustering can hinder the spread of a novel resistant bacterial strain in a host population. We quantify this effect both in the case where resistance pre-exists and in the case where acquiring a new resistance mutation is necessary for the bacteria to spread. We further show that the reduction of spread by clustering can be countered when immune hosts are silent carriers, and are less likely to get treated, and/or have more contacts. We demonstrate the robustness of our findings to including stochastic within-host bacterial growth, a fitness cost of resistance, and its compensation. Our results highlight the importance of interactions between immunity and the spread of antibiotic resistance, and argue in the favor of vaccine-based strategies to combat antibiotic resistance.

\section*{Introduction}

Ever since the discovery of penicillin, every release of a new antibiotic has been followed a few years later by the emergence of bacteria resistant to it \citep{mcclure2014theoretical}. Given the medical importance of antibiotics, resistance is a major public health issue. Treatment can eliminate antibiotic-sensitive bacteria efficiently. But if some bacteria are resistant, then antibiotic treatment will instead select for resistance and increase the proportion of resistant bacteria in the host. Furthermore, the body is home to a numerous and diverse microbiota, mainly in the gut \citep{sender2016revised}, which plays several crucial functional roles \citep{donaldson2016gut,stecher2008role}. Taking an antibiotic treatment against one pathogenic bacterial strain can favor the emergence of drug resistance in other bacteria, in particular in the gut, and these resistant bacteria can then be transmitted, e.g. via the fecal-oral route. This is an important concern because antibiotic use is widespread: for instance, about a quarter of French people are treated with antibiotics every given year \citep{carlet2015rapport,consultations}. Besides, antibiotics are often routinely given to farm animals, and the drug resistance in bacteria they harbor may spread to humans~\citep{dahms2014mini,landers2012review}, though the magnitude of this effect is disputed \citep{harrison2017genomic}. Here, we develop a multiscale model of the interaction between antibiotic use and resistance spread in a host population, focusing on an important aspect of within-host immunity. 

Immunity could interfere with resistance spread in many ways. If the immune system in the gut just massively killed bacteria, it could destabilize the microbiota. Thus, it has to resort to other strategies. Immunoglobulin A (IgA), an antibody isotype which is the main effector of the adaptive immune response secreted in the gut, neither kills its target bacteria nor prevents them from reproducing. It was recently shown in mice that the main effect of IgA is actually to enchain daughter bacteria upon division \citep{Moor2017}. Importantly, clusters of bacteria cannot come close to epithelial cells, which prevents systemic infection and protects the host. Besides, interaction of pathogenic bacteria with epithelial cells can trigger inflammation, which can turn on the bacteria SOS response, increasing horizontal gene transfer between bacteria. Enchained growth thus constitutes a possible mechanism for acquired immunity to dampen horizontal transfer in the gut \citep{diard2017inflammation}. Furthermore, since IgA-mediated clusters of bacteria are mostly clonal, horizontal transfer would most likely occur between very closely related neighboring bacteria, which makes it inefficient at providing new genes. These effects will unequivocally work towards reducing the emergence of antibiotic resistance within the host. In this article, we investigate another, subtler effect. Bacteria being in clonal clusters decreases the effective genetic diversity within the host, and transmitted bacteria are less diverse too. We demonstrate that this can hinder the spread of antibiotic resistance at the scale of the host population. 

New mutations occur upon bacterial replication within a host, but what is crucial for public health is whether these mutant resistant bacteria can spread among the host population. We thus propose a multiscale model, combining within-host dynamics with a stochastic branching process at the between-host scale. Such a description is appropriate at the beginning of epidemic spread, when very few hosts are infected. For instance, a host may be infected with a novel bacterial strain, which can acquire antibiotic resistance by mutation, and which is sufficiently similar to other circulating strains for a portion of the host population to be immune against it. We will consider two types of hosts: ``immune'' hosts secrete IgA antibodies against this novel strain of bacteria in their gut, thus clustering them upon division, while ``naive'' hosts do not.  What is the probability that, starting from one infected individual, this novel bacterial strain invades the host population? First, focusing on the case where the first host is infected with a mix of sensitive and resistant bacteria, we demonstrate that immunity-driven bacteria clustering then decreases the spread probability of the novel strain. We then show that this effect can be reversed if immune and naive hosts have a different number of contacts with other hosts, and a different probability of treatment, which may happen if immune hosts are silent carriers. We further demonstrate the robustness of our findings to more realistic models of the within-host bacterial population dynamics, including mutations, stochasticity, a fitness cost of resistance, and its compensation. Next, we develop analytical approximations of the magnitude of the immunity-driven decrease of spread probability in the case where only sensitive bacteria are initially present, and where the bacterial strain needs to acquire a resistance mutation to spread. Finally, we discuss the implications of our results, notably on the interplay between vaccination and antibiotics.

\section*{Model and methods}

\begin{figure}[h t b]
\centering
\includegraphics[width=0.9\textwidth]{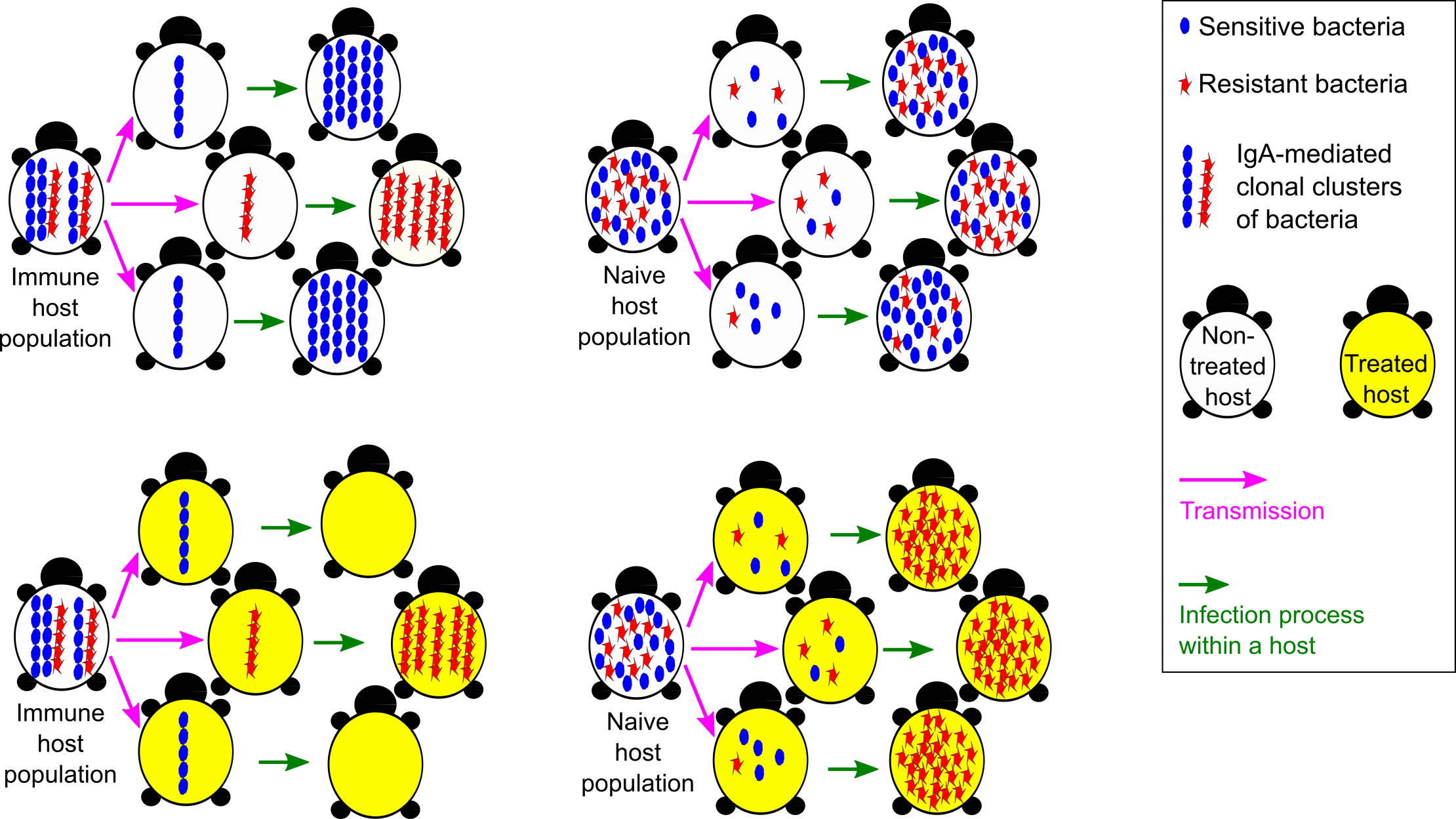}\\
\caption{\label{EffectSketch}\textbf{Sketch illustrating the impact of bacteria clustering on the spread of resistance. } Consider hosts infected by sensitive (blue) and resistant (red) bacteria. Within immune hosts, bacteria form IgA-mediated clonal clusters. This strongly impacts the composition of the set of transmitted bacteria, which is more likely to contain both types of bacteria if the donor host is naive than if it is immune. Antibiotic treatment (yellow) is efficient against sensitive bacteria, but selects for preexisting resistance, which is more likely to exist if the donor host was naive. }
\end{figure}

Here, we describe the multiscale model we developed to demonstrate the impact of antibody-mediated clustering of bacteria on the spread of resistance. Fig.~\ref{EffectSketch} illustrates this effect and the key ingredients of our model.

\subsection*{Within-host dynamics and transmission step}

There is often a typical number of bacteria transmitted from one host to the next for successful infection, called the bottleneck size $N_b$. For instance, $N_b=10^5$ is the typical number of \textit{Salmonella} that starts food poisoning in humans \citep{WHO105}. 
Here, we will assume that infections always start with the same number $N_b$ of infecting bacteria. 

Within the host, the number of bacteria is typically very large. For instance, in a \textit{Salmonella} infection, its density can reach $10^{10}$ bacteria per gram of gut content \citep{Moor2017}. Then, the impact of stochastic fluctuations is likely to be small. Hence, we first use a deterministic model for within-host bacterial population dynamics. We subsequently assess the impact of within-host stochasticity. 

We consider two types of bacteria, a sensitive type, and a resistant type. We first assume that they have the same growth rate, and neglect \textit{de novo} mutations. Then, the proportion of resistant bacteria within a non-treated host remains constant during the infection. Next, we investigate the effects of mutations, of a fitness cost of resistance (see Supporting Information, section \ref{appendix_within-host_growth}), and of the compensation of this cost, for which we use a generalized model with three types of bacteria (see Supporting Information, section~\ref{ThreeTypes}).

When the host is treated with antibiotics, we assume that if it was initially infected only with sensitive bacteria, the treatment is very efficient and kills all bacteria, before resistance appears via mutations, and before any transmission to other hosts. Conversely, if at least one resistant bacteria was present in the inoculum, then the resistant strain takes over. 

In our analytical calculations, we consider that transmitted bacteria are chosen from the donor host using the proportion of resistant bacteria computed from the deterministic model, without correlation between two transmissions from the same host. In particular, we make the approximation that selections of transmitted bacteria are done with replacement. In simulations, we perform selections without replacement, allowing us to check the validity of the approximation.

\subsection*{Impact of clustering on transmission}

Within naive hosts, bacteria remain independent from each other, whereas within immune hosts, they are bound together by the secreted IgA. However, these clusters can break \citep{Tees1993,evans2007forces,mcgrady1987shattering,odell1986flow,bansept2018enchained}. Thus, at the end of the infection, clusters will be of a typical size $N_c$. For simplicity, we will assume that cluster size equals bottleneck size, i.e. $N_c=N_b=N$, meaning that exactly one cluster is transmitted at each infection event. The case where multiple clusters are transmitted to a host ($N_c<N_b$) may also be realistic, but it would be more complex. We wish to focus on the effect of clustering, and the present case $N_c=N_b=N$ will provide an upper bound of the effect of clustering. 

We assume that the concentration of the bacteria studied remains small in the gut, so that the typical encounter time between bacteria or clusters is large. Then, the existing clusters comprise bacteria from the same lineage, since bacteria get enchained by IgA upon replication~\citep{Moor2017}. Then, in the absence of mutations, or when their impact is negligible (e.g. when the initial inoculum already contained non-negligible proportions of both sensitive and resistant bacteria), clusters contain bacteria of the same type, either all sensitive or all resistant. We start by assuming that all clusters are clonal, and then we assess the impact of mixed clusters (see Supporting Information, section \ref{appendix_mixed_cluster}).

We focus on fecal-oral transmissions, and assume that the clusters formed in the gut of a donor immune host are transmitted as clusters. In practice, even if they could break afterwards, bacteria are likely to remain colocalized in the feces. 

\subsection*{Between-host dynamics}

We assume that the number of transmissions to recipient hosts from one donor host is Poisson distributed \citep{Antia2003,Iwasa2004,Andre2005,Orr2008,schreiber2016crossscale} with mean $\lambda_\mathcal{N}$ for naive hosts and $\lambda_\mathcal{I}$ for immune ones. 

We consider that each host has a probability $\omega$ to be immune to the bacteria studied (and thus $1-\omega$ of being naive). The rationale is that we are interested in the spread of a strain at risk to develop resistance, and, while this strain is new, it is similar enough to other strains present in the population for some cross-immunity to exist. We focus on the beginning of the spread of this strain, and thus we neglect the fact that over time, $\omega$ will increase, as infected hosts become immune to this new strain. 

Finally, we assume that hosts receive antibiotic treatment with a probability denoted by $q_\mathcal{N}$ (resp. $q_\mathcal{I}$) for naive (resp. immune) hosts. We initially assume that $q_\mathcal{N}=q_\mathcal{I}$, which is appropriate if antibiotics are given for a reason independent from the bacteria we focus on (e.g. to fight other bacteria, or for growth enhancement in farm animals). We then consider $q_\mathcal{N} \neq q_\mathcal{I}$.

\subsection*{Analytical methods}

Since we focus on the beginning of the spread of a bacterial strain, we use the framework of branching processes \citep{kendall1948generalized,Harris1963,lange2010applied}. We denote by $\wp_i(n_0,n_1,...,n_N)$ the probability for a host initially infected by $i$ resistant and $N-i$ sensitive bacteria to infect $n_0$ hosts with $0$ resistant and $N$ sensitive bacteria, $n_1$ hosts with $1$ resistant and $N-1$ sensitive bacteria, and so on. Let us consider the generating functions $g_i$ of the branching process, for all $i$ between $0$ and $N$:
\begin{equation}
g_i({z_0,z_1,...,z_N})=\sum_{n_0,n_1,...,n_N} \wp_i(n_0,n_1,...,n_N) z_0^{n_0}z_1^{n_1} ... z_N^{n_N}\,,
\label{expr_g_i}
\end{equation}
where each $n_i$ is summed from 0 to infinity.  As we consider no correlation between transmissions, and as the number of transmissions is Poisson distributed, of mean $\tilde{\lambda}$ (with $\tilde{\lambda}=\lambda_\mathcal{N}$ for naive hosts and $\lambda_\mathcal{I}$ for immune ones), we obtain: 
\begin{equation}
g_i({z_0,z_1,...,z_N})=\sum_{k=0}^\infty \frac{\tilde{\lambda}^k e^{-\tilde{\lambda}}}{k!} \left(\sum_{j=0}^N \mathcal{T}_{i,j} z_j\right)^k=\exp\left(-\tilde{\lambda} \sum_{j=0}^N \mathcal{T}_{i,j} (1-z_j)\right) ,
\label{g_i}
\end{equation}
where $\mathcal{T}_{i,j}$ denotes the probability that when a host, initially infected with $i$ resistant bacteria and $N-i$ sensitive ones, infects another host, it transmits to this other host $j$ resistant bacteria and $N-j$ sensitive ones. Note that $\sum_{j=0}^N \mathcal{T}_{i,j}=1$.

Either the new bacterial strain will go extinct, or it will spread to an ever increasing number of hosts, acquiring resistance on the way. The extinction probability $e_i$, starting from one host initially infected with $i$ resistant and $N-i$ sensitive bacteria, is the fixed point of the generating function $g_i$ \citep{Harris1963}. Hence, it satisfies 
\begin{equation}
e_i=\exp\left(-\tilde{\lambda}\sum_{j=0}^N \mathcal{T}_{i,j} \left(1-e_j\right)\right)\,.
\end{equation}
The spread probability is $1-e_i$. 

We will start from these equations for the extinction probabilities, and write them by summing over the cases where hosts are immune or naive, and treated or not. Simplifications are allowed by our assumption that when a host is treated, it does not transmit anything if it was initially infected with no resistant bacteria, and it transmits only resistant bacteria otherwise. Hence, the general equations giving the extinction probabilities read, for all $i$ between 0 and $N$: 
\begin{align}
 e_i =&(1-\omega) (1-q_\mathcal{N}) g_i^\mathcal{N}(e_0,e_1,...,e_N)~ & \} \text{\emph{naive non-treated host}}\nonumber\\
 &+\omega (1-q_\mathcal{I})g_i^\mathcal{I}(e_0,e_1,...,e_N)~ & \} \text{\emph{immune non-treated host}}\nonumber\\
 & +(1-\omega)q_\mathcal{N} \left[\delta_{i0}+(1-\delta_{i0})\exp(-\lambda_\mathcal{N}(1-e_N))\right]~ & \} \text{\emph{naive treated host}}\nonumber\\
 &+\omega q_\mathcal{I} \left[\delta_{i0}+(1-\delta_{i0})\exp(-\lambda_\mathcal{I}(1-e_N))\right]\,,~ & \}\text{\emph{immune treated host}}
 \label{thesystem}
\end{align}
where $\delta_{i0}$ is 1 if $i=0$, and 0 otherwise. Recall that $\omega$ represents the fraction of immune hosts, while $q_\mathcal{N}$ (resp. $q_\mathcal{I}$) is the fraction of treated hosts among the naive (resp. immune) ones. The generating functions $g_i^\mathcal{I}$ and $g_i^\mathcal{N}$, in the immune non-treated and naive non-treated cases, respectively, will be expressed in each case studied.

\subsection*{Numerical methods}

We solve numerically the system of equations Eq.~\eqref{thesystem} giving the values of the extinction probabilities $e_i$. 

We also perform direct numerical simulations of branching processes, both as a test of our analytical descriptions, and because it allows for explicit integration of within-host stochasticity. In our simulations, at each generation of infected hosts, we randomly choose whether each host is immune or not, and treated or not. Next, we simulate within-host growth, either in a deterministic or in a stochastic way. Finally, we perform stochastic transmission of bacteria to the next generation of hosts, with or without clusters, depending whether the transmitting host is immune or not. The process is then iterated. A detailed description of the simulations is provided in the Supporting Information, section~\ref{Simu_SI}. Our code is freely available at \texttt{https://doi.org/10.5281/zenodo.2592323}.

\section*{Results}

\subsection*{Clustering hinders spread in the presence of pre-existing resistance}

What is the impact of IgA-mediated clustering on the probability that a bacterial strain spreads in the host population? To address this question, we first consider the simplest case, with a deterministic within-host growth of the bacterial population, and without mutations or fitness costs of resistance. Then, the proportion of resistant bacteria within a non-treated host remains constant during the infection, equal to $i/N$ if this host was initially infected with $i$ resistant bacteria and $N-i$ sensitive ones. This simplification will allow us to gain insight in the impact of clustering on spread. Next, we will build on this minimal model to address more realistic cases, including within-host growth stochasticity, mutations and fitness costs of resistance. The generating function $g_i^\mathcal{N}$ for naive non-treated hosts is then given by Eq.~\eqref{g_i} with $\tilde{\lambda}=\lambda_\mathcal{N}$, and where the probability $\mathcal{T}_{i,j}$ that a transmission from a host initially infected with $i$ resistant bacteria contains $j$ resistant bacteria follows a binomial law of parameters $N$ and $i/N$:
\begin{equation}
\mathcal{T}_{i,j}=\binom{N}{j}\left(\frac{i}{N}\right)^j\left(\frac{N-i}{N}\right)^{N-j}\,.
\label{binom}
\end{equation}
Besides, the generating function of the process for immune  non-treated hosts reads:
\begin{equation}
g_i^\mathcal{I}(z_0,\dots,z_N)=\exp\left(-\frac{\lambda_\mathcal{I} i}{N}(1-z_N)\right)\exp\left(-\frac{\lambda_\mathcal{I} (N-i)}{N}(1-z_0)\right)\,.
\end{equation}
Using these expressions for the generating functions together with Eq.~\eqref{thesystem} for $0\leq i\leq N$ yields a complete system of equations that allows to solve for the extinction probabilities.

\begin{figure}[h t b]
\centering
\includegraphics[width=0.5\textwidth]{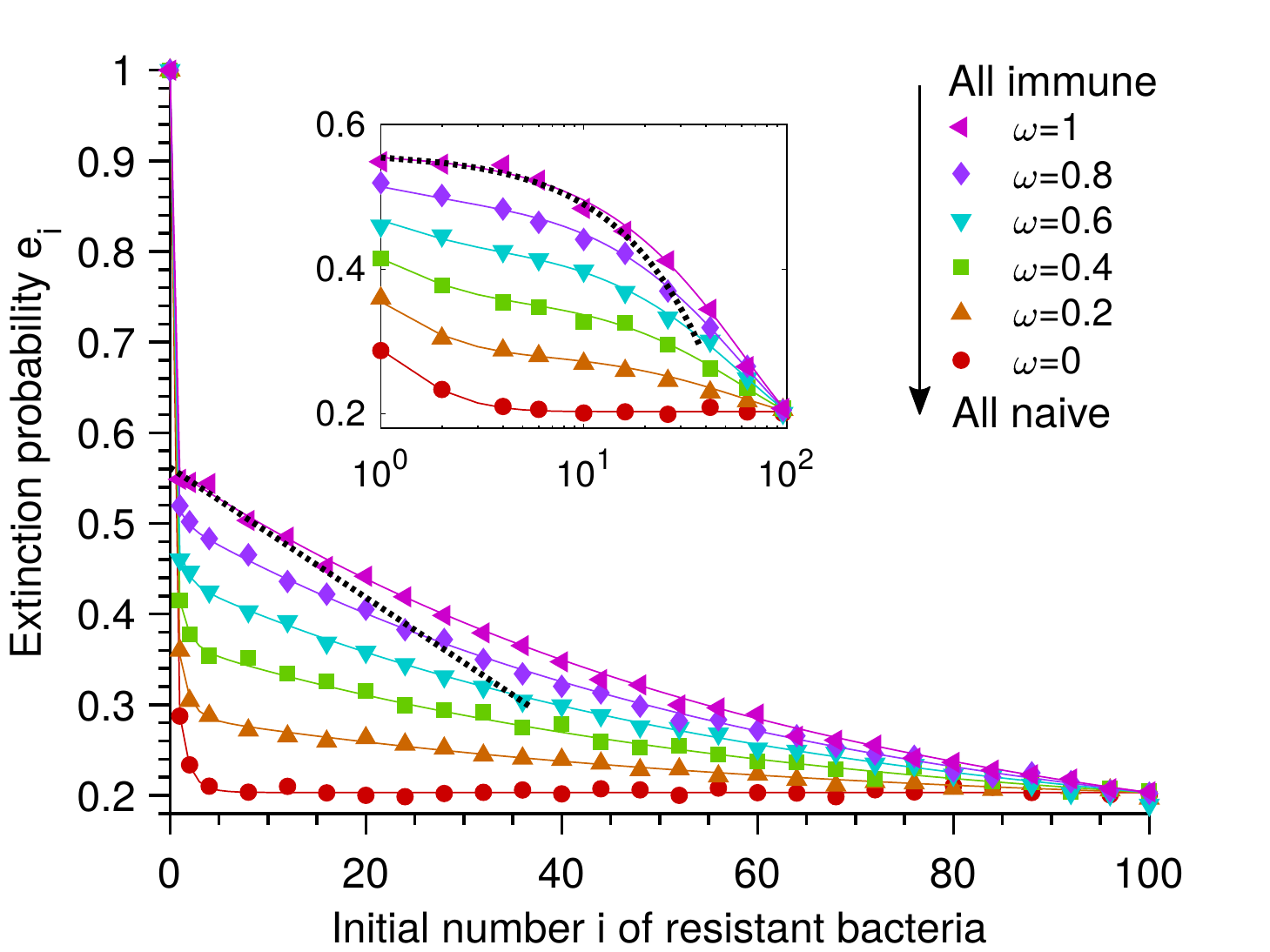}\\
\caption{\label{Fig1}\textbf{Bacterial clustering hinders the spread of a resistant strain in the presence of preexisting resistance.} Extinction probabilities $e_i$ are shown versus initial number $i$ of resistant bacteria in the first infected host, for different fractions $\omega$ of immune individuals in the host population. Naive and immune hosts differ only through bacterial clustering. Each host is treated with probability $q=0.55$ and transmits $N=100$ bacteria to an average of 
$\lambda=2$ other hosts (unless it was infected with no resistant bacteria and treated). Solid lines: numerical resolution of Eq.~\eqref{thesystem}; symbols: simulation results (over $10^4$ realizations). Numerical resolution and simulation are in excellent agreement with each other. Black dotted line: approximation from Eq.~\eqref{appx_eiI} for $\omega=1$ and $0<i\ll N$. Main panel: linear scale; inset: semi-logarithmic scale.}
\end{figure}

A broad range of values for the initial proportion $i/N$ of antibiotic resistance are realistic, from zero or less than a percent to more than half \citep{rhomberg2009summary}. Here, we focus on the case with pre-existing mutations. The case without pre-existing mutations will be addressed in the last part of Results. Fig.~\ref{Fig1} shows that whenever the first host is infected by a mixed inoculum containing both resistant and sensitive bacteria, i.e. $0< i< N$, the extinction probability $e_i$ increases with the proportion $\omega$ of immune hosts, assuming that naive and immune hosts only differ through bacterial clustering. Therefore, bacterial clustering hinders the spread of the strain and of resistance. This effect is strongest for small values of $i$.

Clustering increases extinction probabilities because transmission of resistance is less likely with clustering than without clustering. For a given proportion $i/N$ of resistant bacteria, the probability that a clonal cluster contains no resistant bacteria is $1-\frac{i}{N}$, while a random assortment of $N$ bacteria has a smaller probability $(1-\frac{i}{N})^N$ of containing no resistant bacteria. 

To understand the effect of clustering quantitatively, consider the equations giving the extinction probabilities if all hosts are naive (denoted by $e_i^\mathcal{N}$) or all hosts are immune ($e_i^\mathcal{I}$). If all hosts are naive, Eqs.~\eqref{thesystem} and~\eqref{expr_g_i} yield:
\begin{align}
e_0^\mathcal{N} &=q_\mathcal{N}+(1-q_\mathcal{N}) \exp\left[-\lambda_\mathcal{N}(1-e_0^\mathcal{N})\right], \label{e0b}\\
e_i^\mathcal{N} &=q_\mathcal{N}\exp\left[-\lambda_\mathcal{N}(1-e_N^\mathcal{N})\right]+(1-q_\mathcal{N})\exp\left[-\lambda_\mathcal{N} \sum_{j=0}^N \mathcal{T}_{i,j} (1-e_j^\mathcal{N})\right]\,\,\,\forall i\in[1,N-1], \label{enb}\\
e_N^\mathcal{N} &= \exp\left[-\lambda_\mathcal{N}(1-e_N^\mathcal{N})\right], \label{eNb}
\end{align}
where $\mathcal{T}_{i,j}$ is given in Eq.~\eqref{binom}. Conversely, if all hosts are immune, $e_0^\mathcal{I}$ and $e_N^\mathcal{I}$ are solutions of
\begin{align}
e_0^\mathcal{I} &=q_\mathcal{I}+(1-q_\mathcal{I}) \exp\left[-\lambda_\mathcal{I}(1-e_0^\mathcal{I})\right], \label{e0}\\
e_N^\mathcal{I} &= \exp\left[-\lambda_\mathcal{I}(1-e_N^\mathcal{I})\right], \label{eN}
\end{align}
and $e_i^\mathcal{I}$ can be obtained from $e_0^\mathcal{I}$ and $e_N^\mathcal{I}$ via
\begin{equation}
e_i^\mathcal{I}=q_\mathcal{I} e_N^\mathcal{I}+\left[(1-q_\mathcal{I})e_N^\mathcal{I}\right]^\frac{i}{N}(e_0^\mathcal{I}-q_\mathcal{I})^{1-\frac{i}{N}}\,. \label{en}
\end{equation}
These equations show that if naive and immune hosts differ only through bacterial clustering, i.e. $\lambda_\mathcal{N}=\lambda_\mathcal{I}=\lambda$ and $q_\mathcal{N}=q_\mathcal{I}=q$, then $e_N^\mathcal{N}=e_N^\mathcal{I}= e_N$ and $e_0^\mathcal{N}=e_0^\mathcal{I}= e_0$. Hence, clustering is irrelevant if all bacteria are identical. In addition, $e_0>e_N$ if $q>0$ and $\lambda>1$ (see Eqs.~\eqref{e0b} and \eqref{eNb}): in the presence of treatment, resistance decreases the extinction probability. 

Consider now a first host infected with both resistant and sensitive bacteria ($0<i<N$). If this host is treated, its infection becomes fully resistant, leading to an extinction probability $e_N^\mathcal{N}$ or $e_N^\mathcal{I}$, depending whether the host population is fully naive or fully immune. In a fully naive host population, if the first host is not treated, its probability of not transmitting resistance to a given new host is $(1-\frac{i}{N})^N$. For $N\gg 1$ and $0<i\ll N$, $(1-\frac{i}{N})^N\approx \exp(-i)$, which is smaller than 5\% for $i>2$: then, transmission of resistance is almost certain, and 
\begin{equation}
e_i^\mathcal{N}\approx e_N^\mathcal{N}.
\label{almost_eN}
\end{equation}
In a fully immune population, if the first host is not treated, and if  $(1-q_\mathcal{I})\lambda_\mathcal{I}<1$, spread is possible only if it transmits a resistant cluster (probability $i/N$) to one of the recipient hosts (which are on average $\lambda_\mathcal{I}$), and this leads to spread (probability $1-e_N^\mathcal{I}$), yielding an extinction probability $\approx 1-(1-e_N^\mathcal{I}) \lambda_\mathcal{I} i/N$. This approximation neglects the case where multiple recipient hosts receive resistant bacteria, which is appropriate if $\lambda_\mathcal{I} i/N \ll 1$. Hence, if $0<i< N$ and $\lambda_\mathcal{I} i/N \ll 1$ and $(1-q_\mathcal{I})\lambda_\mathcal{I}<1$,
\begin{equation}
e_i^\mathcal{I}\approx q_\mathcal{I} e_N^\mathcal{I}+(1-q_\mathcal{I})\left(1-(1-e_N^\mathcal{I})\lambda_\mathcal{I}\frac{i}{N} \right),
\label{appx_eiI}
\end{equation}
which is consistent with Eq.~\eqref{en} for $i/N\rightarrow 0$ and $e_0^\mathcal{I}=1$. In Fig.~\ref{Fig1}, for $\omega=0$, we observe an early plateau as $i$ increases, as predicted by Eq.~\eqref{almost_eN}, and for $\omega=1$, our complete results are well approximated by Eq.~\eqref{appx_eiI} when $0<i\ll N$.

Importantly, in the present case where naive and immune hosts differ only through bacterial clustering, i.e. $\lambda_\mathcal{N}=\lambda_\mathcal{I}=\lambda$ and $q_\mathcal{N}=q_\mathcal{I}=q$, we have $e_N^\mathcal{N} = e_N^\mathcal{I}$, and thus Eqs.~\eqref{almost_eN} and \eqref{appx_eiI} yield $ 1- e_i^\mathcal{I} \approx (q+(1-q) \lambda i/N) (1- e_i^\mathcal{N})$. Hence, immunity reduces the spread probability by a factor up to $1/q$.

\subsection*{The reduction of spread can be countered by silent carrier effects}

So far, we assumed that immune and naive hosts differ only through IgA-mediated bacterial clustering. However, considering pathogenic bacteria, immune hosts may feel less sick than naive ones, because clustering prevents direct interaction with epithelial cells, and thus systemic infection. Immune hosts may still shed bacteria: they are silent carriers. Sick naive hosts may have fewer contacts than immune ones (e.g. if sick hosts stay at home), i.e. $\lambda_\mathcal{N}<\lambda_\mathcal{I}$. Note that one might also imagine the opposite effect, e.g. because the infection is cleared faster in immune hosts, but $\lambda_\mathcal{N}>\lambda_\mathcal{I}$ would reinforce the reduction of resistance spread due to immunity. Hence, we focus on $\lambda_\mathcal{N}<\lambda_\mathcal{I}$, which might counter this effect. In addition, if antibiotic treatment is given in response to infection by the bacterial strain considered, silent carrier immune hosts might be treated less often, i.e. $q_\mathcal{I}\leq q_\mathcal{N}$. Immune hosts then become reservoirs of sensitive bacteria, which can either favor the emergence of resistance by enabling more spread, or hinder it, because a reduced use of antibiotics decreases the competitive advantage of resistant bacteria. Here, we investigate the interplay of clustering with these silent carrier effects. 

\begin{figure}[h t b]
\centering
\includegraphics[width=\linewidth]{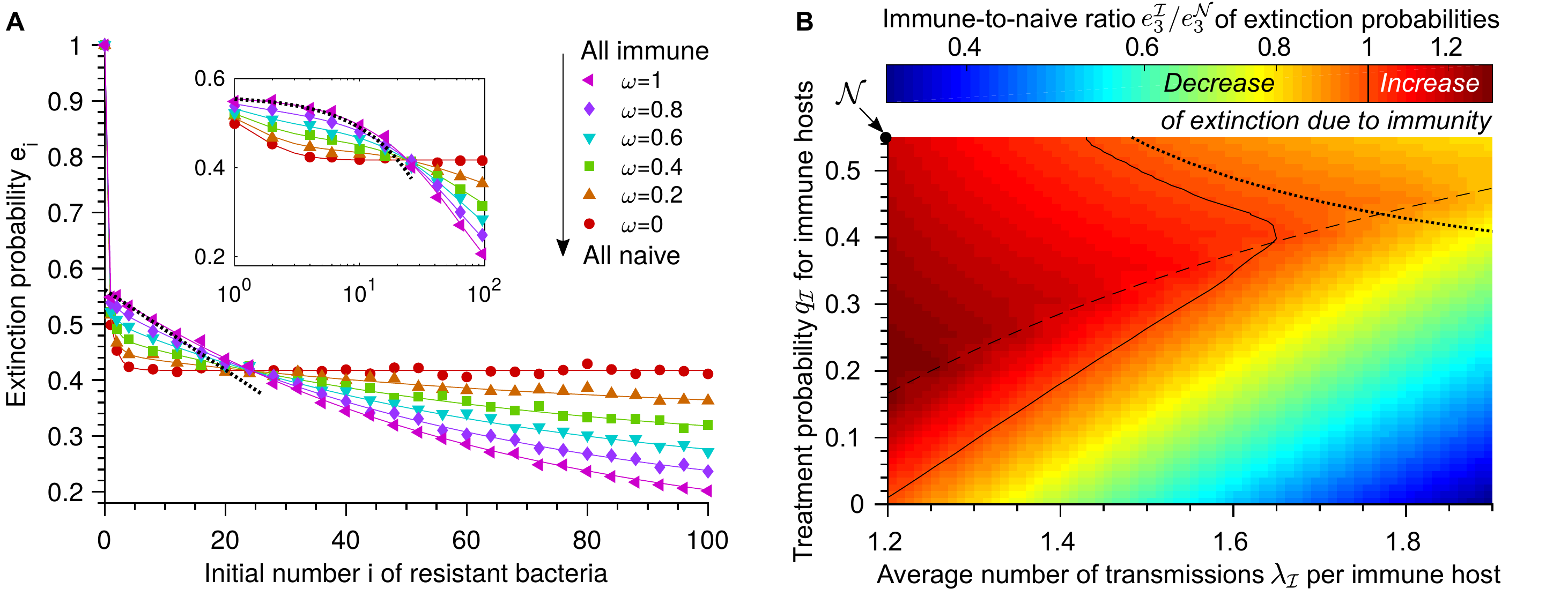}
\caption{\label{FigSilent}\textbf{Silent carrier effects can counter the clustering-driven reduction of spread.} \textbf{A:} Extinction probabilities $e_i$ versus initial number $i$ of resistant bacteria in the first infected host, for different fractions $\omega$ of immune individuals in the host population. Parameters are the same as in Fig.~\ref{Fig1}, except that each naive (resp. immune) host transmits to an average of $\lambda_\mathcal{N}=1.5$ (resp. $\lambda_\mathcal{I}=2$) other hosts. Solid lines: numerical resolution of Eq.~\eqref{thesystem}; symbols: simulation results (over $10^4$ realizations). Black dotted line: approximation from Eq.~\eqref{appx_eiI} for $\omega=1$ and $0<i\ll N$. \textbf{B:} Heatmap: the ratio $e_3^\mathcal{I}/e_3^\mathcal{N}$ of the extinction probabilities in a fully immune versus a fully naive population, starting from $i=3$, is shown for $\lambda_\mathcal{I}\geq \lambda_\mathcal{N}=1.2$ and $q_\mathcal{I}\leq q_\mathcal{N}=0.55$. Values ($\lambda_\mathcal{N}$, $q_\mathcal{N}$) for naive hosts are indicated by a dot labeled ``$\mathcal{N}$''. Parameters are the same as in Fig.~\ref{Fig1}, except $\lambda_\mathcal{I}$, $\lambda_\mathcal{N}$ and $q_\mathcal{I}$. Solid curve: $e_3^\mathcal{I}/e_3^\mathcal{N}=1$. Dashed curve: $\lambda_\mathcal{I}(1-q_\mathcal{I})=1$. Dotted curve: $q_\mathcal{I}=(1-e_N^\mathcal{N})/(1-e_N^\mathcal{I})$ (Eq.~\eqref{cdnSilent}). Heatmap interpolated from numerical resolutions of Eq.~\eqref{thesystem}; logarithmic color scale.}
\end{figure}

Fig.~\ref{FigSilent}A shows the extinction probability $e_i$ as a function of the initial number $i$ of resistant bacteria in the first infected host, as in Fig.~\ref{Fig1}, but for $\lambda_\mathcal{N}<\lambda_\mathcal{I}$. Other parameters are the same, and we remain in the regime where extinction is not certain with only resistant bacteria, but certain without resistance, i.e. $1<\lambda_\mathcal{N}<\lambda_\mathcal{I}$, while $\lambda_\mathcal{N} (1-q_\mathcal{N})<1$ and $\lambda_\mathcal{I} (1-q_\mathcal{I})<1$. Fig.~\ref{FigSilent}A shows that when $0< i\ll N$, the extinction probability increases with the proportion $\omega$ of immune hosts, as in Fig.~\ref{Fig1}. But strikingly, this effect is reversed for larger values of $i$. Therefore, the reduction of resistance spread by bacterial clustering can be countered by silent carrier effects. 

Let us analyze this trade-off by comparing a fully immune population with a fully naive one. If $1<\lambda_\mathcal{N}<\lambda_\mathcal{I}$, Eqs.~\eqref{eNb} and \eqref{eN} yield $e_N^\mathcal{I} < e_N^\mathcal{N}<1$. If in addition $q_\mathcal{I}\leq q_\mathcal{N}$, then $\lambda_\mathcal{I}(1-q_\mathcal{I}) > \lambda_\mathcal{N} (1-q_\mathcal{N})$, and either $e_0^\mathcal{I}=e_0^\mathcal{N}=1$ if $\lambda_\mathcal{I}(1-q_\mathcal{I})<1$, or $e_0^\mathcal{I}< e_0^\mathcal{N}$ if $\lambda_\mathcal{I}(1-q_\mathcal{I})>1$ (see Eqs. \eqref{e0b} and \eqref{e0}). Hence, if all bacteria are resistant, immunity actually favors the spread, because of the associated increased transmission. If all bacteria are sensitive, the same conclusion holds if $\lambda_\mathcal{I}(1-q_\mathcal{I})>1$. Let us now consider intermediate initial numbers of resistant bacteria. If $0<\lambda_\mathcal{I} i/N \ll 1$ and $\lambda_\mathcal{I}(1-q_\mathcal{I})<1$, then $e_i^\mathcal{I}$ satisfies Eq.~\eqref{appx_eiI}, which simplifies into $e_i^\mathcal{I}\approx 1-q_\mathcal{I}(1-e_N^\mathcal{I})$ to zeroth order in $\lambda_\mathcal{I} i/N$. Besides, $e_i^\mathcal{N}$ satisfies Eq.~\eqref{almost_eN} for $2<i\ll N$. Employing these approximations, the condition $e_i^\mathcal{I}>e_i^\mathcal{N}$ becomes 
\begin{equation}
q_\mathcal{I}<\frac{1-e_N^\mathcal{N}}{1-e_N^\mathcal{I}}\,.
\label{cdnSilent}
\end{equation}
If $1<\lambda_\mathcal{N}<\lambda_\mathcal{I}$, then $e_N^\mathcal{I} < e_N^\mathcal{N}<1$, so Eq.~\eqref{cdnSilent} can hold or not depending on $q_\mathcal{I}$, which means that immunity can hinder the spread of resistance or not. Fig.~\ref{FigSilent}A is in the regime where immunity increases extinction probabilities for $2<i\ll N$ (Eq.~\eqref{cdnSilent} holds), but decreases them for $i=N$. 

Fig.~\ref{FigSilent}B shows a heatmap of $e_3^\mathcal{I}/e_3^\mathcal{N}$, for $\lambda_\mathcal{I}$ and $q_\mathcal{I}$ satisfying $1<\lambda_\mathcal{N}\leq\lambda_\mathcal{I}$ and $q_\mathcal{I}\leq q_\mathcal{N}$. We observe that increasing $\lambda_\mathcal{I}$ decreases $e_3^\mathcal{I}/e_3^\mathcal{N}$, but that the impact of decreasing $q_\mathcal{I}$ is subtler, since it can both enable more spread, and decrease the competitive advantage of resistant bacteria. Moreover, as predicted, if $\lambda_\mathcal{I}(1-q_\mathcal{I})<1$ and Eq.~\eqref{cdnSilent} both hold, then $e_3^\mathcal{I}/e_3^\mathcal{N}>1$, and immunity hinders the spread of resistance. Fig.~\ref{FigSilent}B also features a region where $\lambda_\mathcal{I}(1-q_\mathcal{I})>1$ (thus $e_0^\mathcal{I}< e_0^\mathcal{N}$) but $e_3^\mathcal{I}/e_3^\mathcal{N}>1$. Elsewhere, the reduction of resistance spread by immunity is countered by silent carrier effects, and $e_3^\mathcal{I}/e_3^\mathcal{N}<1$.

\subsection*{Results are robust to including mutations, within-host growth stochasticity, and a cost of resistance}

So far, we considered the simple case where the within-host growth simply preserves the proportion of resistant bacteria. We now include more realistic within-host population dynamics. 

First, what is the impact of mutations on the spread of resistance in the presence of immunity? Typical mutation rates for bacteria are $\sim 10^{-10}-4\times 10^{-9}$ per base pair per replication \citep{lynch2010evolution}. There are often several mutations conferring resistance, yielding an effective total mutation probability $\mu_1 \sim 10^{-10} - 10^{-6}$ from sensitive to resistant at each replication \citep{zurWiesch11}. The back-mutation probability $\mu_{-1}$ from resistant to sensitive is generally substantially smaller, as the exact same mutation has to be reverted \citep{Levin00}. Hence, while developing a complete model that includes back-mutations, we will often present cases where they are neglected. The total number of bacterial generations $G$ within a host can vary. For instance, in experimental infection of mice by \textit{Salmonella} starting at different inoculum sizes, $G$ is typically 10 (inoculum of $10^7$ bacteria) to 35 (inoculum of $10^3$ bacteria) after 24h~\citep{Moor2017}. Assuming deterministic exponential within-host growth, we employ differential equations to compute the fraction of resistant bacteria at the end of the incubation time, and we write the corresponding generating functions (see Supporting Information, section \ref{appendix_within-host_growth}). 

\begin{figure}[h t b]
\centering
\includegraphics[width=\linewidth]{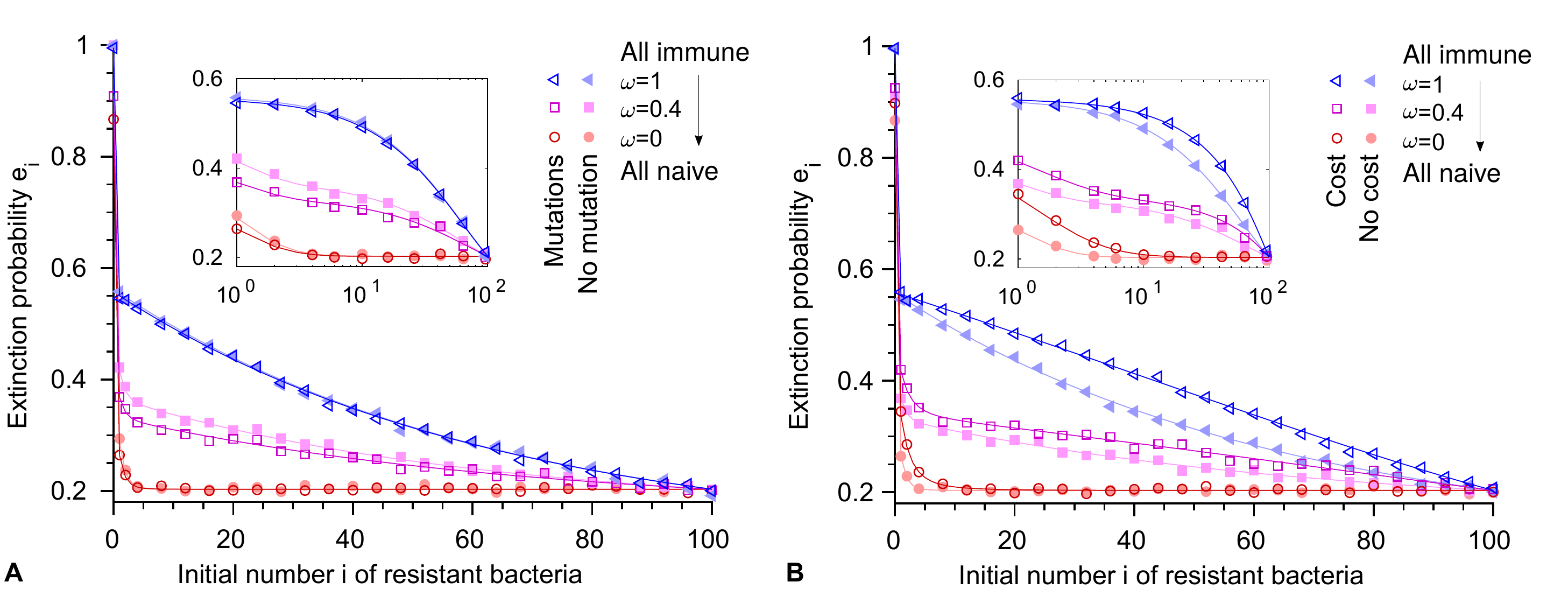}
\caption{\label{FusionFig}\textbf{Impact of mutations and of a fitness cost of resistance.} Extinction probabilities $e_i$ versus initial number $i$ of resistant bacteria in the first infected host, for different fractions $\omega$ of immune individuals in the host population. \textbf{A:} Results with mutations are shown in dark shades. Results without mutation (identical to Fig.~\ref{Fig1}) are shown in light shades for comparison. \textbf{B:} Results with a fitness cost $\delta=0.1$ of resistance are shown in dark shades. Results without fitness cost (same as dark-shaded results in panel A) are shown in light shades for comparison. In both panels, within-host evolution is deterministic, $\mu_1=7 \times 10^{-5}$ and $\mu_{-1}=0$, and incubation time is $G=10$ generations. As in Fig.~\ref{Fig1}, $N=100$, $\lambda=2$ and $q=0.55$. Solid lines: numerical resolution of Eq.~\eqref{thesystem}; symbols: simulation results (over $10^4$ realizations).}
\end{figure}

Fig.~\ref{FusionFig}A compares results obtained with and without mutations. Overall, accounting for mutations from sensitive to resistant slightly decreases the extinction probabilities $e_i$, especially for small values of $i$, including $i=0$, where extinction is no longer certain. Indeed, mutations tend to increase the fraction of resistant bacteria, with a stronger effect if this fraction is initially small. Fig.~\ref{FusionFig}A further shows that the impact of mutations is small for a fully immune population. Then, for the small proportions of resistant bacteria such that mutations matter, the probability that a non-treated host transmits resistant bacteria is small anyway because of clustering. Interestingly, Fig.~\ref{FusionFig}A shows that the impact of mutations is stronger for an intermediate fraction $\omega$ of immune hosts than for a fully naive host population. This is due to cross-transmission events: for small $i$, if an immune host infects a naive host, it very likely transmits only sensitive bacteria. But if the naive recipient host is not treated, mutations can induce \textit{de novo} resistance, which is then likely to be transmitted. Hence, the propagation of resistance is less likely to be stopped by immune hosts with mutations than without mutations.

So far, we considered deterministic within-host growth. To assess the impact of within-host growth stochasticity on the spread of resistance in the presence of immunity, we performed simulations with stochastic exponential within-host growth but without bacterial death or loss (see Supporting Information, section~\ref{Simu_SI}). In the absence of mutations (Fig.~\ref{FigMutS}), the only difference with our deterministic model is the stochasticity of growth, while with mutations (Fig.~\ref{FigMut2}), an additional difference is that mutations are treated in a stochastic way, which is more realistic since mutations are rare events.  We found that stochasticity matters most for naive hosts and small initial numbers $i$ of resistant bacteria, but overall, its impact is small, which validates our use of a deterministic model (Supporting Information, section \ref{simuSIresults}). We further showed that including small loss rates of bacteria does not affect our conclusions (Supporting Information, section~\ref{appendix_loss}), and that our assumption of neglecting mixed clusters is valid when the inoculum is mixed (Fig.~\ref{FigMixedClust}).

Mutations conferring resistance often carry a fitness cost. Denoting by $f$ the growth rate of sensitive bacteria without antibiotics, and by $f(1-\delta)$ that of resistant bacteria, typical values for $\delta$ range from 0.005 to 0.3 \citep{Levin00,Andersson10,zurWiesch11,deSousa15} (but can be even larger~\citep{Paulander07}). Here, we assume that this fitness cost only affects within-host growth, and not transmissibility \citep{miran_cross,schreiber2016crossscale}. Fig.~\ref{FusionFig}B shows that extinction probabilities are increased by a fitness cost, taken equal to 0.1. Indeed, the fraction of resistant bacteria after deterministic within-host growth, starting from a given inoculum, is smaller with a cost than without one. In a fully immune population, the cost has less impact for small values of the initial number $i$ of resistant bacteria, because transmission of resistance is then very unlikely anyway (Eq.~\eqref{appx_eiI}). In contrast, in a fully naive population, the fitness cost has a substantial impact for small $i$, but very little impact for larger $i$, because transmission of resistance is then very likely (Eq.~\eqref{almost_eN}).

The fitness cost of many resistance mutations can be compensated by subsequent mutations \citep{Levin00,Schrag97,maisnier2004adaptation,Paulander07,deSousa15,Hughes15,Marrec18}. Hence, we generalized our model to include compensation (see Supporting Information, section~\ref{ThreeTypes}). While compensation would determine long-term survival of resistance if the treatment was stopped or became less frequent, our results (Figs.~\ref{FigCostS} and \ref{FigCost2}) show that it does not have a major impact on the initial steps of the propagation studied here.

Overall, our finding that clustering hinders the spread of resistance is robust to including mutations, within-host stochasticity, as well as a fitness cost and its compensation.

\subsection*{Spread probability without pre-existing resistance}

We now develop analytical approximations to better characterize the impact of antibody-mediated clustering on the spread of resistance, starting without pre-existing resistant mutants ($i=0$). Here, we take into account mutations and the fitness cost of resistance, but not its compensation. We use the deterministic description of the within-host growth, and take $\lambda_\mathcal{N}=\lambda_\mathcal{I}=\lambda$, as well as  $q_\mathcal{N}=q_\mathcal{I}=q$. 

Let us focus on the case where the bacterial strain would certainly go extinct without mutations, i.e. $\lambda (1-q) <1$ (otherwise, immunity has little impact on its spread). Then, the spread probability, starting from a host infected with only sensitive bacteria, is proportional to $\mu_1$, and thus very small. Hence, we assume that at most one event leads to spread. In the absence of mutations, the mean number of infected hosts is $1+\lambda (1-q)+ (\lambda (1-q))^2+...=1/(1-\lambda (1-q))$, both for naive and immune hosts. We will express the probability that each transmission from an infected host contains resistant bacteria, and that this leads to spread, comparing fully immune and fully naive host populations.

So far, we have considered that antibody-mediated bacterial clusters are either fully sensitive, or fully resistant, which is appropriate when most mutants are descendants of a pre-existing mutant transmitted to the host, as confirmed by Fig.~\ref{FigMixedClust}. However, in a host initially infected with only sensitive bacteria, mixed clusters may matter. A simplified view of the process of cluster growth and breaking is that clusters grow in linear chains and break in two once a certain size is attained, then grow and break again, and so on. As bacteria are enchained upon division, the subclusters formed upon breaking comprise closely related bacteria. Assume that the maximal cluster size is $2^g=N$: it is attained in $g\leq G$ generations, where $G$ is the number of generations within a host. When a mutation occurs, the cluster comprises mixed bacterial types until $g$ generations after (see Fig.~\ref{schema_clusters_mut_break}, Supporting Information section \ref{appendix_mixed_cluster}). In the limit of a small mutation rate, the final proportion of fully mutant clusters corresponds to the proportion of mutant bacteria at generation $G-g$, seeding the final clusters. Meanwhile, neglecting selection during cluster growth, which is valid for $g\delta \ll 1$, the final proportion of mixed clusters is equal to the probability that a mutation occurs during the final cluster growth, i.e. $2 \mu_1 (2^g-1)=2 \mu_1 (N-1)$. Let us now consider in more detail two extreme regimes, depending on the number of generations $G$ within a host.

\paragraph{Small $G$. } 

Assume that $\delta G \ll 1$: selection weakly impacts the proportions of bacterial types, and thus, the final proportion of fully mutant clusters is $\approx \mu_1(G-g)$. Assume also that $(G-g) \ll 2 (N-1)$: starting with a fully sensitive infection, most transmissions of mutant bacteria happen through mixed clusters. 

In this regime, if at least one resistant mutant is transmitted, the probability of extinction is similar for an immune and a naive host population (see Supporting Information section \ref{smallGapproxsection}). Besides, in a fully naive population, the probability that a host infected with only sensitive bacteria transmits at least one resistant bacteria is $\approx\mu_1 G N$ (the proportion of resistant bacteria multiplied by the bottleneck size). Conversely, in a fully immune population, this probability is $\approx 2 \mu_1 (N-1)$ (the proportion of mixed clusters). Thus, the ratio of spread probabilities reads
\begin{equation}
\mathcal{R}=\frac{1-e_0^\mathcal{N}}{1-e_0^\mathcal{I}}\approx \frac{G}{2}\,.
\label{eqR1}
\end{equation}

\paragraph{Large $G$.} 

Assume that $\delta G > 1$: the proportion of resistant bacteria in infections started with a mixed inoculum end up close to the mutation-selection balance. Assume also that the number of mixed clusters, $2 \mu_1 (N-1)$, is small compared to the number of fully mutant clusters, $ (1-\exp(-(G-g)\delta))\mu_1/\delta$ (see Supporting Information section \ref{proportionmutants}): this condition can be expressed as $2 N \delta \ll 1$. Assume that $G\gg g$: the proportion of resistant bacteria at $G-g$ is well approximated by its value $r_0$ at $G$. Finally, assume that $N \mu_{-1} (1+\delta)^G \ll 1$, i.e. back-mutations can be neglected: hosts infected with only resistant bacteria generally transmit only resistant bacteria. Thus, a host infected with only resistant bacteria leads to an spread probability $1-e_N$ similar for naive and immune hosts. 

In this regime, how can spread occur, starting from a host infected with only sensitive bacteria? In a fully naive host population, spread occurs if the following three events happen. First, a resistant bacterium is transmitted (probability $\approx r_0 N$, the final proportion of resistant bacteria multiplied by the bottleneck size). Second, the recipient host is treated (probability $q$; else, the final proportion of resistant bacteria is close to the mutation-selection balance, leading to negligible spread probability). Third, the resulting fully resistant infection spreads with probability $1-e_N$. Hence, the spread probability per host infected with only sensitive bacteria is $\approx r_0 N q (1-e_N)$. Besides, in a fully immune host population, spread occurs if a cluster of resistant bacteria is transmitted (probability $\approx r_0$), and emerges (probability $1-e_N$). Thus, the spread probability per host infected with only sensitive bacteria is $\approx r_0 (1-e_N)$. The ratio of spread probabilities then reads
\begin{equation}
\mathcal{R}=\frac{1-e_0^\mathcal{N}}{1-e_0^\mathcal{I}}\approx qN\,.
\label{eqR2}
\end{equation}

\begin{figure}[h!]
\centering
     \includegraphics[width=\linewidth]{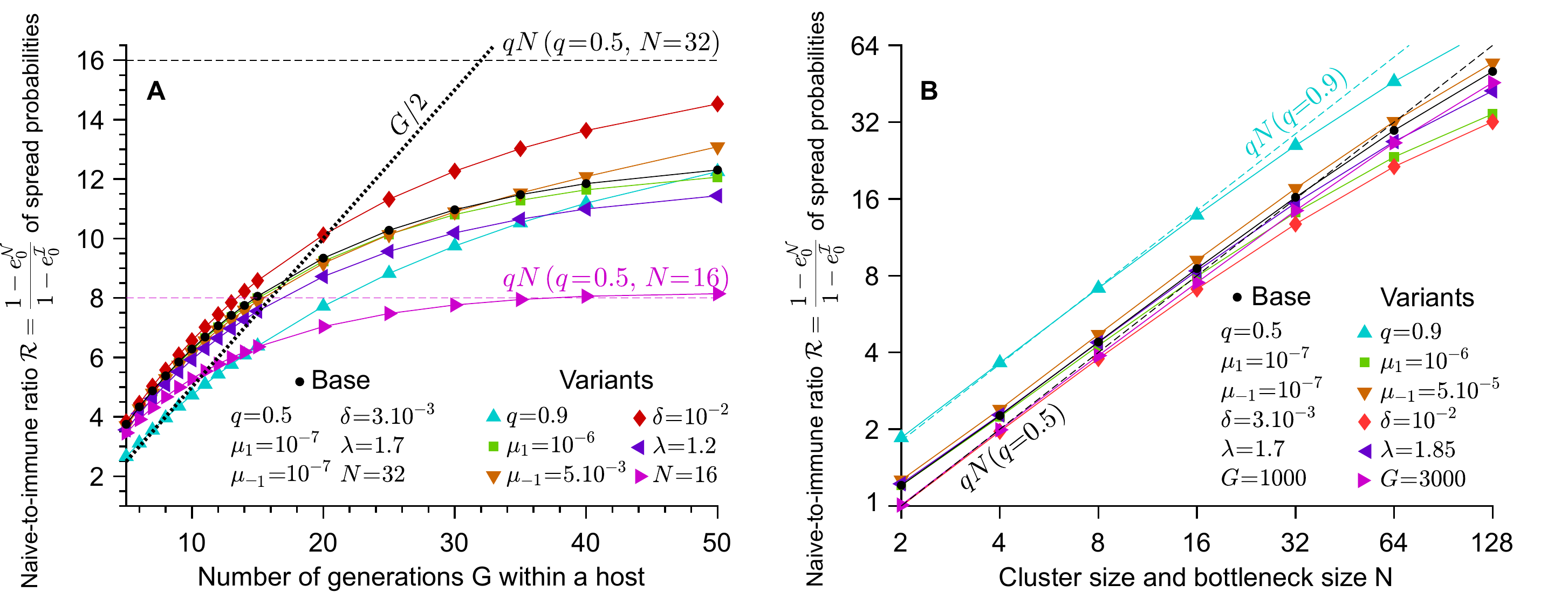}
\caption{\textbf{Ratio $\mathcal{R}$ of the spread probabilities in a fully naive population relative to a fully immune population, without pre-existing resistance.} \textbf{A:} $\mathcal{R}$ versus the number of within-host generations $G$. \textbf{B:} $\mathcal{R}$ versus the cluster and bottleneck size $N$. Markers joined with lines (both panels): exact values of $\mathcal{R}$ obtained by numerical resolution of the full system of equations yielding extinction probabilities (see Supporting Information section \ref{FullsystemSection}). Results with base parameters are plotted in black, and variants showing the impact of changing each single parameter are plotted in colors.  Dashed lines: analytical approximation Eq.~\eqref{eqR2} of $\mathcal{R}$ for large $G$. Black dotted line (panel A): analytical approximation Eq.~\eqref{eqR1} of $\mathcal{R}$ for small $G$. This figure shows that the simple expressions in Eqs.~\eqref{eqR1} and~\eqref{eqR2} are robust and capture well the dependence of $\mathcal{R}$ on the various parameters. }\label{approxGfig}
\end{figure}

Fig.~\ref{approxGfig} demonstrates that the simple expressions in Eqs.~\eqref{eqR1} and~\eqref{eqR2} are valid. In conclusion, without pre-existing resistance, and when mutations are necessary for spread, immunity can substantially decrease the spread probability.

\section*{Discussion}

At the scale of a population of hosts, bacterial adaptation via evolution, for instance the spread of antibiotic resistant mutations, could be affected by host immunity. In particular, IgA, an antibody isotype which is the main effector of the adaptive immune response in the gut, could have important impacts in the case of fecal-oral transmission, although it neither kills bacteria nor prevents them from reproducing. It was shown recently \citep{Moor2017} that its main effect is to enchain bacteria: bacteria recognized by IgA remain stuck together after replication. This clustering may reduce inflammation and horizontal gene transfer \citep{diard2017inflammation}. Here, we shed light on another original effect of clustering: the reduced bacterial diversity resulting from this clustering could hinder the spread of adaptive mutations in a population of hosts, even if the total number of bacteria in a host remains the same. In practice, this direct effect of clustering could work in synergy with the reduction of horizontal gene transfer \citep{diard2017inflammation} to reduce the spread of antibiotic resistance.

To assess how antibody-mediated bacteria clustering impacts the spread of antibiotic resistance, we developed a multi-scale model, with a deterministic description of within-host evolution, and a stochastic description of between-host transmission, assuming a probability $q$ for host treatment, and a fixed bottleneck size $N$ at transmission (taken equal to the bacterial cluster size; this case gives an upper bound to the effect of clustering). We showed that if the first host is infected by a mix of sensitive and resistant bacteria, immunity decreases the spread probability of the bacterial strain by a factor up to $1/q$, assuming that infection in naive and immune hosts differs only through bacterial clustering. We demonstrated that this effect can be countered when an immune host is a silent carrier, and is less likely to get treated, and/or has more contacts. Our results proved robust to the introduction of mutations and stochasticity in the within-host dynamics. We also showed that if resistance comes at a cost, extinction probabilities are increased, while compensation of this cost has little impact, except for long incubation periods. However, compensation would matter for the long-term maintenance of resistance, e.g. if treatment was stopped. Finally, we showed that if only sensitive bacteria are initially present, and if the bacterial strain needs to acquire a resistance mutation to spread, immunity decreases the spread probability by a factor approximately $G/2$ when the number $G$ of within-host generations is small, and approximately $qN$ when $G$ is large. 

Immunity hinders the spread of drug resistance because immune hosts mostly transmit bacteria of the same type, due to the antibody-mediated clonal clusters they carry. The disadvantage of a higher variance (here, in the number of resistant bacteria transmitted) is a classic theme in population genetics~\citep{gillespie1974natural,frank1990evolution}, and also appears in analyses of viral adaptation~\citep{virus_life}. For a given average number of transmitted resistant bacteria, the proportion of transmissions comprising at least one resistant bacteria is smaller for immune donor hosts than for naive ones. In a sense, transmissions from immune hosts ``put all their eggs in one basket'', and prevent bacteria from hedging their bets.

Microbial populations infecting a host often grow to very large numbers, so that even with realistic small mutation rates, many mutants may be produced. Single point mutations can suffice to confer resistance \citep{Andersson10}, implying that resistance-carrying mutants might exist in most infected hosts. Nevertheless, the spread of drug resistance in a host population typically takes years~\citep{mcclure2014theoretical}. One reason is that only a small proportion of the large within-host microbial population is transmitted to the next host. This bottleneck decreases diversity, and most mutants are not transmitted, because they are rare. Because of this sampling effect, the immunity-mediated clustering of clonal bacteria can hinder the spread of antibiotic resistance, and more generally of any type of adaptive mutations. Note that using a deterministic description of within-host growth, and excluding bacterial death or loss, transmission from naive hosts involving one cluster is effectively equivalent to a bottleneck size of one single bacterium. However, for such small bottlenecks, stochasticity would significantly matter. Reducing the bottleneck size has been shown to generally increase the extinction probability in other contexts, but sometimes has the opposite effect~\citep{schreiber2016crossscale,leclair2018impact}. The details of transmission play an important part in this effect. Our multi-scale description integrating realistic aspects of the within-host dynamics, in particular immunity, thus allowed us to gain important insight into the spread of a pathogen in a host population. While our study was motivated by IgA-induced bacterial clustering, our results would extend to other mechanisms yielding clonal clusters. There are other host effectors besides IgA that cluster bacteria together, for instance neutrophil extracellular traps \citep{fournier2012role}.

Our work highlights the importance of interactions between immunity and the spread of antibiotic resistance. In particular, antibiotic treatment and vaccines inducing IgA production might have non-trivial interplays. In practice, vaccines protect the host, but may or may not reduce shedding of bacteria~\citep{desin2013salmonella,sharma2018shedding}. Reduction of shedding clearly hinders pathogen propagation.  However, our results demonstrate that even a vaccine that does not reduce shedding can protect the host population. Indeed, it can increase the efficiency of antibiotic treatment, by hindering the spread of resistance in the host population. Our results thus constitute an additional argument in the favor of vaccine-based strategies to combat antibiotic resistance, which are the focus of renewed interest~\citep{AMR,Lipsitch16,Jansen18}.

\subsection*{Acknowledgments}
All authors thank Raphaël Voituriez both for stimulating scientific discussions and for unfailing financial support. FB and LM acknowledge funding by graduate fellowships from EDPIF.



\newpage

\singlespacing

\setcounter{page}{1}

\section*{\Huge{Supporting Information}}

\def\theequation{S\arabic{equation}}
\setcounter{equation}{0}
\def\thefigure{S\arabic{figure}}
\setcounter{figure}{0}

\tableofcontents

\newpage
\section{Table of the symbols used }
\label{tabsymbols}

\begin{center}
 \begin{tabular}{|p{0.9cm}|p{12cm}|}
 \hline
 \multicolumn{2}{|c|}{\textbf{Parameters specific to the infection}}\\
 \hline
  $G$ & Duration in number of replications of the within-host infection\\
 \hline
 $N_b$ & Typical bottleneck size, i.e. the number of bacteria seeding the infection in a new host\\
  \hline
  $N_c$ & $=2^g$ Maximum cluster size (when they reach it, they break in half before the next replication). In general we take $N_c=N_b=N=2^g$\\
  \hline
 \multicolumn{2}{|c|}{\textbf{Within-host dynamics}}\\
 \hline
$S(t)$  & Number of sensitive bacteria within a specific host at time $t$	\\
  \hline
$R(t)$  & Number of resistant bacteria within a specific host at time $t$	\\
 \hline
 $f$  & Division rate of sensitive bacteria	\\
 \hline
   $\delta$  & Fitness cost of resistance, such that $f(1-\delta)$ is the division rate of resistant bacteria	\\
 \hline
 $\mu_1$  & Probability for each of the two daughter bacteria of a sensitive bacterium to become resistant because of a mutation during replication	\\
 \hline
 $\mu_{-1}$  & Probability for each of the two daughter bacteria of a resistant bacterium to become sensitive because of a mutation during replication\\
  \hline
  $\mu_2$  & In the model with three bacterial types, probability for each of the two daughter bacteria of a resistant bacterium to become resistant-compensated because of a mutation during replication	\\
 \hline
 $r_i$ & Proportion at transmission of resistant bacteria within a host that was initially infected with $i$ resistant and $N-i$ sensitive\\
  \hline
 \multicolumn{2}{|c|}{\textbf{Immune vs. Naive hosts}}\\
 \hline
  $\omega$ & Proportion of immune individuals in the host population\\
  \hline
$q_\mathcal{N}$ & Proportion of naive individuals in the host population who are antibiotic-treated\\
  \hline
  $q_\mathcal{I}$ & Proportion of immune individuals  in the host population who are antibiotic-treated\\
  \hline
  $\lambda_\mathcal{N}$ & Mean number of contacts a naive host transmits the infection to\\
  \hline
  $\lambda_\mathcal{I}$ & Mean number of contacts an immune host transmits the infection to\\
  \hline 
  \multicolumn{2}{|c|}{\textbf{Systems of equations}}\\
 \hline
  $e_i$ & Probability of extinction for an infection that was seeded in patient zero by $i$ resistant bacteria and $N-i$ sensitive bacteria \\
  \hline
  $\mathcal{T}_{i,j}$ & Probability that when a host, initially infected with $i$ resistant and $N-i$ sensitive bacteria, infects another host, it transmits $j$ resistant bacteria and $N-j$ sensitive ones.\\
  \hline
 
 \end{tabular}
\end{center}

\newpage


\section{Within-host growth equations and generating function}\label{appendix_within-host_growth}

Below, we present the deterministic differential equations on the numbers of sensitive and resistant bacteria within a non-treated infected host including mutations and a fitness cost of resistance. We assume exponential growth, and we simplify using $\mu_1,\mu_{-1} \ll\delta  \ll 1$. We denote by $G$ the total number of replications within a host, corresponding to the incubation time. We also denote by $r_i$ the proportion of resistant bacteria in a host at the end of incubation, for a host that was initially infected with $i$ resistant bacteria and $N_b-i$ sensitive ones (recall that each host is infected by a total of $N_b$ bacteria). Here $N_b$ is the bottleneck size, in practice we generally take it equal to the cluster size $N_c$ and denote both by $N$.

\subsection{Discrete vs. continuous time representation}

Let us first consider the case without mutations, with  $S$ the number of sensitive bacteria, $R$ the number of resistant bacteria, and $\delta $ the fitness cost of resistance. If there are $G$ generations for the sensitive strain, there are $G(1-\delta)$ generations for the resistant strain. Denoting by $f$ the growth rate of sensitive bacteria, the ordinary differential equations governing the growth of the bacterial population read: 
\begin{equation}
 \frac{dS}{dt}=fS
\end{equation}
\begin{equation}
 \frac{dR}{dt}=f(1-\delta)R
\end{equation}
The solutions of these equations are $S(t)=S_0 \exp(ft)$, $R(t)=R_0\exp(f(1-\delta)t)$. 
If we start from one bacteria of each type, after a time $\tau$ corresponding to $G$ generations of the sensitive bacteria, there are $2^{G}=\exp(f\tau)$ sensitive bacteria, and $2^{G(1-\delta)}=\exp(f(1-\delta)\tau)$ resistant bacteria. Thus:
\begin{equation}\label{corrGtau}
 G \log(2)=f\tau.
\end{equation}

Next, let us consider the case with mutations. When a sensitive bacteria divides, each of the daughter bacteria has a probability $\mu_1$ to be mutant (and thus to have become resistant in our model). When a resistant bacterium divides, each of the daughter bacteria has a probability $\mu_{-1}$ to have mutated (and thus to have become sensitive in our model). Let us denote $\tilde{\mu}_1$ and $\tilde{\mu}_{-1}$ the mutation rates for the system of differential equations, such that: 
\begin{equation}
 \frac{dS}{dt}=f(1-\tilde{\mu}_1)S+\tilde{\mu}_{-1} f (1-\delta) R
 \label{ode1}
\end{equation}
\begin{equation}\label{eqRdebut}
 \frac{dR}{dt}=f(1-\delta)(1-\tilde{\mu}_{-1})R+\tilde{\mu}_1 f S
\end{equation}

We then look for the relation between $\mu_i$ (discrete model) and $\tilde{\mu_i}$ (continuous model). The accumulation of mutants in the early dynamics has to be the same.  
Let us start from sensitive bacteria only, neglect back mutations, and take the limit of very small $\delta$. Then, when considering a bacteria after $G$ generations, there have been $G$ opportunities for mutation, i.e. the proportion of resistant bacteria is $G \mu_1$. In our continuous description, if there were $S_0$ sensitive bacteria (and no resistant bacteria) at $t=0$, and still neglecting back mutations, $S(t)=S_0 \exp(f(1-\tilde{\mu}_1)t)$, and, replacing $S(t)$ by this expression in \eqref{eqRdebut}, and solving for $R(t)$ with $R(0)=0$:
\begin{equation}
R(t)=S_0 \tilde{\mu}_1 \exp(f(1-\tilde{\mu}_1)t) \frac{\exp(f (\tilde{\mu}_1+ \tilde{\mu}_{-1} (-1+\delta ) - \delta ) t)-1 }{\tilde{\mu}_1 +
  \tilde{\mu}_{-1} (-1+\delta ) - \delta }. 
\end{equation}
In the limit of small $t$, 
\begin{equation}
R(t)\approx S_0 \tilde{\mu}_1 ft\exp(f(1-\tilde{\mu}_1)t) 
\end{equation}
 and the proportion of resistant bacteria then reads: 
\begin{equation}
r(t)=\frac{R(t)}{R(t)+S(t)}\approx \frac{R(t)}{S(t)}= \tilde{\mu}_1 ft
\end{equation} 
Thus $G \mu_1=\tilde{\mu}_1 f\tau =\tilde{\mu}_1G \log(2)$ (where we used Eq.~\eqref{corrGtau}), and consequently we have to take $\tilde{\mu}_1=\mu_1/\log(2)$ for consistency.

\subsection{Resolution of the continuous-time system}

When the host is not treated, the within-host dynamics can be complex. The growth could be limited by some carrying capacity and taken as logistic, there could be a loss term, etc. As we want to calculate the proportions of sensitive and resistant bacteria, the following equations will give similar results as equations with a carrying capacity:
\begin{equation}\label{eqr}
 \frac{d S}{d t_g}= (1-\mu_1/\log(2)) S+(1-\delta) \mu_{-1} R /\log(2),
\end{equation}
\begin{equation}\label{eqs}
 \frac{d R}{d t_g}= (1-\delta)(1-\mu_{-1}/\log(2)) R+\mu_1 S/\log(2),
\end{equation}
with $t_g$ the time in numbers of generations. 
The aim is to obtain the proportion of sensitive and resistant bacteria at the end of the infection within a host, depending on the composition of the inoculum. 

The total number of replications within a host  $G$ can vary. The typical minimal doubling time for bacteria is half an  hour \citep{mason1935comparison}. Bacterial carriage can last several days or even more, but when close to carrying capacity, the growth rate decreases. As a portion of the bacteria will be lost in feces, there will be ongoing replication, though at a lower rate. Thus $G$ can take a wide range of values. For instance, in experimental infection of mice by \textit{Salmonella} starting at different inoculum sizes, the number of replications is typically 10 (inoculum of $10^7$ bacteria) to 35 (inoculum of $10^3$ bacteria) after 24h  \citep{Moor2017}. 

Solving Eqs.~\eqref{eqr} and \eqref{eqs} with the initial conditions $S(0)=N-i$ and $R(0)=i$, we find for all $i$ between $0$ and $N$ the following exact expression for the proportion $r_i=\frac{R}{R+S}$ of resistant bacteria after $G$ generations, knowing that the infection was seeded with $i$ resistant and $N-i$ sensitive bacteria at time $t=0$:
\begin{equation}
 r_i = \frac{(2^{\Delta G} -1) ( 2 \mu_1 N + i (-\mu_1 - \mu_{-1} - \delta \log(2) + \mu_{-1} \delta )) + i \Delta\log(2) (2^{\Delta G}+1 )}{N((2^{\Delta G} -1) (\mu_1 + \mu_{-1} + \delta \log(2) - 2 i \delta  \log(2)/N - \mu_{-1} \delta ) + 
     \log(2)\Delta (2^{\Delta G}+1 ))}
\end{equation}
with 
\begin{equation}\textstyle
\Delta= \sqrt{\delta ^2 \left(1 - \frac{\mu_{-1}}{\log(2)}\right)^2+ 2 \delta  \left(-\frac{\mu_1}{\log(2)} + \frac{\mu_{-1}}{\log(2)} - \frac{\mu_1 \mu_{-1}}{\log^2(2)} -\frac{\mu_{-1}^2}{\log^2(2)} \right)+ \frac{(\mu_1 + \mu_{-1})^2}{\log^2(2)} } .
\label{Delta}\end{equation}
\vspace{0.5cm}
Let us look at particular cases in more detail.

\subparagraph{Case with only sensitive bacteria initially:} In this case, as $1 \gg \delta  \gg \mu_1,\mu_{-1}$, the final proportion of resistant bacteria can be approximated to:
\begin{equation}\label{eqp0}
 r_0 \approx \mu_1  \frac{1-2^{-\delta G} }{\delta  \log(2)}. 
\end{equation}

\subparagraph{Case with mixed inoculum:} Let us distinguish two cases: 
\begin{itemize}
\item If $\delta  G \ll 1$, when starting with both strains, their relative proportion will have little time to change. 
If there is one resistant bacteria initially, then let us neglect mutations in both ways, as the mutation rate is small, and then the final proportion of resistant bacteria is: 
\begin{equation}\label{eqp1}
 r_1\approx \frac{1}{1+(N_b-1)(1+ G \log(2) \delta )}
\end{equation} 
\item If $\delta  G \gg 1$, then if there was at least one resistant and one sensitive bacteria initially, the mutation-selection balance is reached within the infected host. The final proportion of resistant bacteria  $r_{MSB}$ is obtained by writing the differential equation on $\frac{R}{R+S}$ and looking for its equilibrium. In the limits we are considering, $r_{MSB}$ tends to $\mu_1/(\delta  \log(2))$.
\end{itemize}

\subparagraph{Case starting from resistant bacteria only:} In this case, the final proportion of resistant bacteria will be: 
\begin{equation}\label{eqpNfull}
 r_N =1- \frac{2 (2^{\Delta G}-1) \mu_{-1} (1-\delta)}{( 2^{\Delta G}-1) ( \mu_1 + \mu_{-1} - \delta  \log(2) - \delta \mu_{-1})  + (1 +2^{\Delta G}) \Delta \log(2)}
\end{equation}
with $\Delta$ defined in Eq. \eqref{Delta}.
For $G$ not too large, since $1 \gg \delta \gg \mu_1,\mu_{-1}$, then $\Delta \approx \delta$. Consequently:
\begin{equation}\label{eqpN}
 r_N \approx 1- \frac{(2^{\delta G}-1)\mu_{-1} (1-\delta)}{2^{\delta G} \mu_{-1} (1-\delta) +\delta  \log(2)}. 
\end{equation}

\subsection{Expression of the generating functions using within-host growth results}

\label{FullsystemSection}

In all the expressions below, we consider a host which is not treated. 

\subsubsection{Naive hosts}

In naive hosts initially infected with $i$ mutant bacteria, when there is transmission, the probability to transmit $j$ resistant bacteria and $N-j$ sensitive ones reads: 
\begin{equation}
 \mathcal{T}_{i,j}=\binom{N}{j} r_i^j (1-r_i)^{N-j}. 
\end{equation}
Then, for all $i$ between 0 and $N$: 
\begin{equation}
 g_i^\mathcal{N}(e_0,e_1,...,e_N)= \exp\left(- \lambda_\mathcal{N} \left(\sum_{j=0}^N \binom{N}{j} r_i^j (1-r_i)^{N-j} (1-e_j) \right) \right). 
\end{equation}

\subsubsection{Immune hosts}

In immune hosts, let us address the question of whether the clusters comprise bacteria of a single type, or are mixed. As mentioned in the main text and detailed in Supporting Information section \ref{appendix_mixed_cluster}, a simplified view of the process of cluster growth and breaking is that clusters break in two once a certain size is attained, then grow and break again, and so on. As bacteria are enchained upon division, the subclusters formed upon breaking comprise closely related bacteria. Assume that the maximal cluster size is $2^g=N$: it is attained in $g\leq G$ generations, where $G$ is the number of generations within a host.

\paragraph{Mixed inoculum.}

As explained in section \ref{appendix_mixed_cluster}, the clusters are made of daughter cells enchained together. Thus, in the absence of mutations, clusters will comprise bacteria of one type only. As mutation rates are very small, when the initial inoculum is mixed, the proportion of mixed clusters is very small compared to the proportion of clusters comprising bacteria of one type only. Therefore, we neglect mixed clusters when the initial inoculum is mixed. 
Thus for all $i$ between $1$ and $N-1$, 
\begin{equation}
 g_i^\mathcal{I}(e_0,e_1,...,e_N)= \exp\left(- \lambda_\mathcal{I} \left((1-r_i) (1-e_0)+r_i (1-e_N) \right)\right). 
\end{equation}

\paragraph{Inoculum of only one bacterial type.}

To be fully mutant at generation $G$, the clusters need to be seeded by a bacteria which was mutant at generation $G-g$. This happens with probability $r'_0$ when the inoculum is fully sensitive and $1-r'_N$ when the inoculum is fully resistant (with $r'_i$ denoting the proportion of resistant bacteria at generation $G-g$ while $r_i$ is the one at generation $G$). 

For mixed clusters (see section \ref{conclusion_mixed}), we assume  $\delta g \ll 1$, so that differences in growth time between the different types of clusters can be neglected. 
A cluster at generation $G$ was founded by one bacteria at generation $G-g$. A mutation occurs at its first replication with probability $2\mu$, resulting in a cluster of size $2^g$ containing $2^{g-1}$ mutants. The probability for mutation will be $2^2\mu$ at the next replication, and so on. Thus, the probability for a cluster to include one mutant is $2^g \mu = N \mu $, that to include 2 mutants is $2^{g-1} \mu = N \mu / 2$, that to include 4 mutants is  $2^{g-2} \mu = N \mu / 2^2$, ... , and finally, that to include $2^{g-1}$ mutants is $ 2 \mu$. This yields: 

\begin{equation}
 g_0^\mathcal{I}(e_0,e_1,...,e_N)= \exp\left[- \lambda_\mathcal{I} \left((1-2 \mu_1 (N-1) - r'_0 ) (1-e_0)+\sum_{j=0}^{g-1} N \mu_1 \frac{1-e_{2^j}}{2^j} +r'_0 (1-e_N)\right)\right]
\end{equation}

\small

\begin{equation}
 g_N^\mathcal{I}(e_0,e_1,...,e_N)= \exp\left[- \lambda_\mathcal{I} \left(( r'_N-2 \mu_{-1} (N-1) ) (1-e_N) +\sum_{j=0}^{g-1} N \mu_{-1} \frac{1-e_{N-2^j}}{2^j} +(1-r'_N) (1-e_0)  \right) \right]
\end{equation}
\normalsize

\section{Dynamics of clusters including mixed clusters, and probability to transmit at least one mutant}
\label{appendix_mixed_cluster}

\subsection{A simple view of cluster growth}

\begin{figure}[h t b]
\centering
\includegraphics[width=0.75\textwidth]{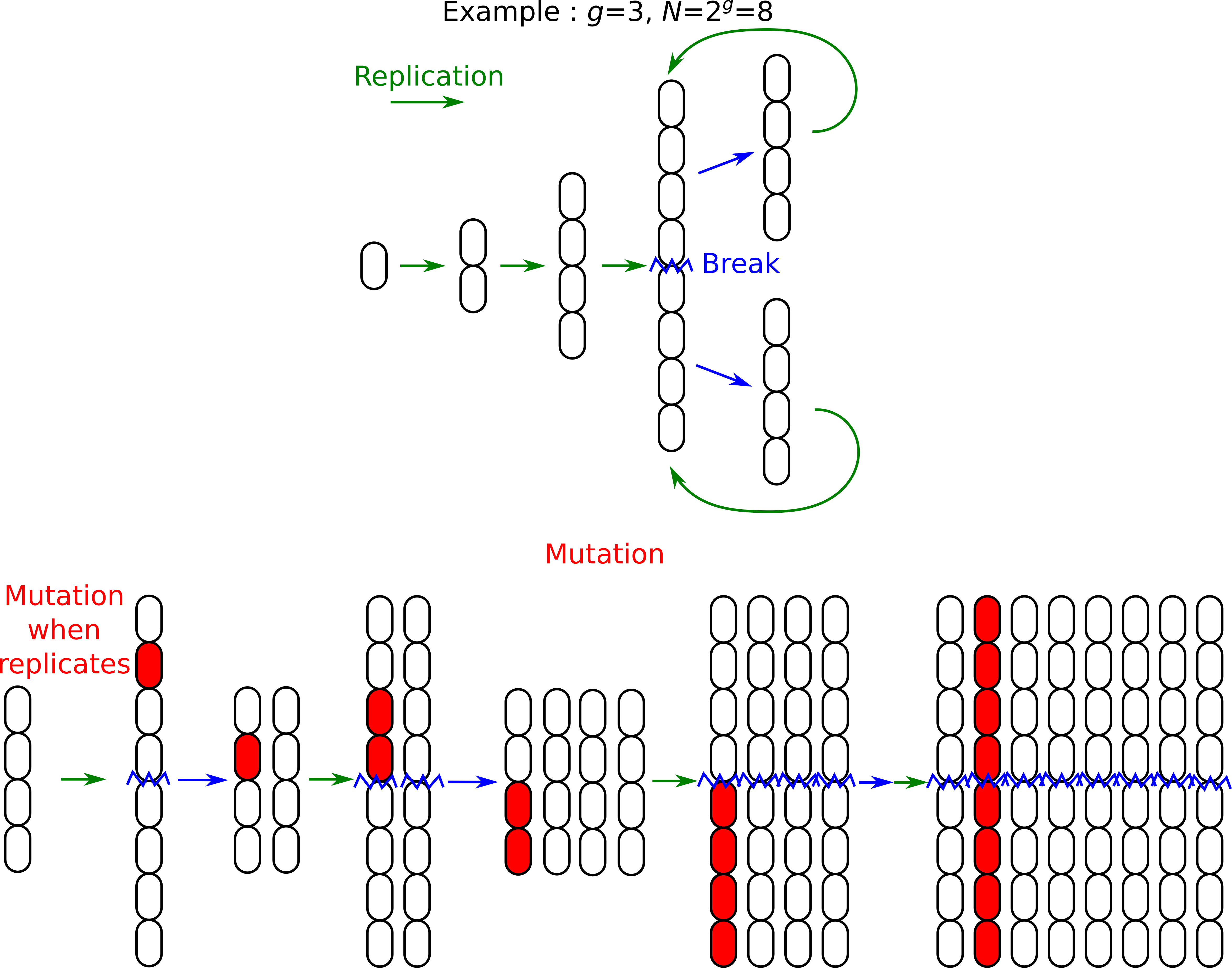}\\
\caption{\label{schema_clusters_mut_break}\textbf{Schematic showing the outcome of mutations in clusters. } Represented here is the case of simple linear clusters. For more complex clusters, it will remain true that closely related bacteria are located close to one another, since daughter bacteria remain bound together after replication.}
\end{figure}

As mentioned in the main text, a simplified view of the process of cluster growth and breaking is that clusters break in two once a certain size is attained, then grow and break again, and so on. As daughter bacteria are physically close in the clusters formed through IgA-mediated enchained growth, the subclusters formed after a larger cluster breaks will comprise closely related bacteria. Let us assume that a host infection lasts for $G$ generations total, and that the maximal cluster size is $2^g=N_c$ (see schematic in Fig.~\ref{schema_clusters_mut_break}): this size is attained in $g$ generations ($2^g=N_c$), with $g\leq G$.  We here assume that $N_c$, the cluster size, is equal to $N_b$, the bottleneck size, and we denote them both as $N$. When a mutation occurs, the cluster comprises mixed bacterial types until $g$ generations after (see Fig.~\ref{schema_clusters_mut_break}). In the limit of a small mutation rate, the final proportion of fully mutant clusters corresponds to the proportion of mutant bacteria at generation $G-g$, seeding the final clusters.

\subsection{Probability to transmit at least one mutant}
\label{probaAuMoinsUnMut}

Here, we consider a host, initially infected with only one type of bacteria, with a mutation rate $\mu$, and no fitness cost for the mutation. We look at the probability to transmit at least one mutant, and ask how our simple model could differ from a more realistic one. Note that here, we do not consider hosts infected with a mixed inoculum, because in this case, as loss of bacteria is rare at the beginning of the infection, and as the population of bacteria grows to large numbers in the infected host, the number of both sensitive and resistant bacteria is then expected to be large, so that mutations and fluctuations are not expected to play a big role. 

In our model, we effectively assume that the proportion of resistant bacteria at the end of the infection can be considered equal to its mean expected value (the average being over several realizations of the within-host dynamics), which is given by the deterministic ordinary differential equations. However, fluctuations could be important. For instance, if the infection starts from a small number of bacteria of the same type, a mutant appearing during the first replication will give rise to an important share of mutants in the population at the end of the infection. As explained in section~\ref{FullsystemSection}, we also usually assume that in immune individuals, except if the infection starts with only sensitive bacteria, clusters comprise only one type of bacteria, either all sensitive or all resistant. In this section, we explore how realistic these assumptions are. 

Let us consider two extreme cases of population dynamics, starting with only sensitive bacteria: 
\begin{itemize}
 \item Exponential growth: starting from size $N$, bacteria divide $G$ times, leading to a final population size $N 2^G$. At each replication, each daughter bacteria has a probability $\mu$ of mutating. This is the case considered in the main text and above.
 \item Constant population size: let us assume that very quickly, the population grows to size $N_{pop}$ (we neglect mutations in this phase, and consider than we start with $N_{pop}$ sensitive bacteria, and then at each of the $G$ steps, the bacteria all replicate (with a probability of mutation $\mu$ for each daughter bacteria), and then half of the bacteria are removed, so the population size remains constant. 
\end{itemize}
In both cases, on average, taking bacteria at the end of the infection, they went through $G$ replications from the start of the infection, so they have a probability $\mu G$ of being mutant. When transmitting $N$ bacteria (and assuming that $N$ is very small compared to the final population size, and that $N \mu G \ll 1$), the probability to transmit one mutant is $\approx N \mu G$ from a naive host. From an immune host, we assume that mutants are transmitted only in fully mutant clusters, with probability $\mu G$. Here we will study the validity of these assumptions.

Henceforth, we will consider the case of a very small mutation rate $\mu$ so that at most one mutation occurs during the infection of a host. 

\subsubsection{Exponential growth}
\label{probaAuMoinsUnMutExp}

Recall that in this regime, starting from size $N$, the bacteria divide $G$ times, leading to a final population size $N 2^G$. At each replication, each daughter bacteria has a probability $\mu$ of mutating. 

\paragraph{Naive host.}

Consider the lineage of one bacteria: it involves $G$ steps of replication. At step $j$, this bacteria has $2^j$ descendants, and there is a probability $\mu 2^j$ that one mutation occurs at this step, in which case $2^{G-j}$ bacteria will carry the mutation in the final population. This will correspond to a proportion $\rho_j=2^{G-j}/(2^G N)= 1/(N 2^j)$ in the final population. 

Assuming that $2^G \gg 1$, we can neglect the difference between taking a sample with or without replacement, so the probability for a transmission from a naive host to contain at least one mutant will be $1-(1-\rho_j)^{N}$ if the mutation occurs at step $j$ (probability $1-\rho_j$ for one bacteria to be of the initial type, probability $(1-\rho_j)^{N}$ that all bacteria chosen are of the initial type). So, multiplying by the initial $N$ bacteria, and summing over the $G$ replication steps, the probability that a transmission from a naive host includes at least one mutant is: 
\begin{equation}\label{p1expnaive}
 m_1^\mathcal{N}=\sum_{j=1}^G N \mu 2^j \left(1-\left(1- \frac{1}{N 2^j} \right)^{N} \right)
\end{equation}

In the limit of $j$ large, $(1- \frac{1}{N 2^j})^{N} \approx 1-1/2^j$, so Eq.~\eqref{p1expnaive} yields $m_{1,exp}^\mathcal{N} \approx N \mu G$. Hence, the result from our simple model employing ordinary differential equations is recovered. We know that this result is an upper bound of the real value: an early mutation will lead to a higher proportion of mutants in an individual, and thus the probability that several mutants are transmitted at the same time will be higher. But, because when we average over all possible transmissions, the mean number of mutants does not change, a higher probability of transmitting several mutants at a time leads to a lower probability of transmitting at least one mutant.  

In the limit of $N$ large enough for $1-\exp(N \log(1-1/(N 2^j))) \geq 1- \exp(- 1/2^j)$ to hold (note that this does not require $N$ to be extremely large as $N 2^j\gg 1$ is sufficient), Eq.~\eqref{p1expnaive} yields:  
\begin{equation}
 \begin{split}
  m_1^\mathcal{N} &\geq \sum_{j=1}^G N \mu 2^j \left(1-\exp(-1/2^j) \right) \\
 & \geq \sum_{j=1}^G N \mu 2^j \left(1/2^j -1/(2^{2j+1})\right)\\
 & = \sum_{j=1}^G N \mu  \left(1 -1/(2^{j+1})\right)\\
 & = N\mu \left( G- \frac{1}{2^2}\sum_{j=1}^G \frac{1}{2^{j-1}}\right)\\
 &=N \mu \left(G - \frac{1}{2}\left(1- \frac{1}{2^G} \right) \right)\\
 \end{split}
\end{equation}
Thus, we have shown that:
\begin{equation}\label{boundsnaive}
  N \mu G \geq m_{1,exp}^\mathcal{N} \geq N \mu (G-1/2). 
\end{equation}

\paragraph{Immune host.}
In the immune case, we assume that one cluster is transmitted. Let us estimate the probability that the cluster transmitted is fully mutant ($m_N^\mathcal{I}$) and the probability that the cluster transmitted is mixed ($m_{mixed}^\mathcal{I}$). 
If a mutation occurs at a step $j$ (probability $N \mu 2^j$) between the first step and the $(G-g)^{th}$ step, then there will be $2^{G-g-j}$ mutant bacteria at the $(G-g)^{th}$ step, yielding $2^{G-g-j}$ fully mutant clusters at the final $G^{th}$ step. They will then be in proportion $1/(N 2^j)$ among the $N 2^{G-g}$ clusters. 
Thus: 
\begin{equation}\label{pnexpimmune}
 m_N^\mathcal{I}= \sum_{j=1}^{G-g} N \mu 2^j \frac{1}{N 2^j} = \mu (G-g). 
\end{equation}
If a mutation occurs at step $j$ between the $(G-g+1)$-th step and the $G$-th step, then it will give one mixed cluster. Thus: 
\begin{equation}\label{pmixedexpimmune}
m_{mixed,exp}^\mathcal{I}=  \sum_{j=G-g+1}^{G} N \mu 2^j \frac{1}{N 2^{G-g}} =  \sum_{j=1}^{g} \mu 2^j = \mu 2 (2^g-1) = 2 \mu (N -1)
\end{equation}

\paragraph{Conclusion.}\label{conclusion_mixed}

Interestingly, when the host is naive, the result is very close to the mean-field case. Indeed, we showed that the probability to transmit at least one mutant is bounded between $2N \mu (G-1/2)$ and $2 N \mu G$ (the mean field result) (see Eq.~\eqref{boundsnaive}). The total probability for a transmission from an immune host to include at least one mutant is $m_{tot}^\mathcal{I}= m_N^\mathcal{I} +  m_{mixed}^\mathcal{I} = \mu (G-g+2(2^g-1))=  \mu (G-\log_2(N)+2(N-1))$ (see Eqs. \eqref{pnexpimmune} and \eqref{pmixedexpimmune}). If $N \gg G$, then it gives $m_{tot,exp}^\mathcal{I} \approx 2 N \mu$, which is $G/2$ times smaller than for the naive case. However in this case most transmissions will be of mixed clusters rather than fully mutant clusters.  If $N \ll G$, then $m_{tot,exp}^\mathcal{I} \approx  \mu G$, which is $N$ times smaller than for a naive host, and will be mostly of fully mutant clusters.

\subsubsection{Constant population size}
\label{cstpopsz}

Recall that in this regime, we assume that very quickly, the population grows to size $N_{pop}$ (we neglect the mutations at the beginning, so that we consider that we start with $N_{pop}$ bacteria of the initial type), and then at each of the $G$ steps, the bacteria all replicate (with a probability of mutation $\mu$ for each daughter bacteria), then half of the bacteria are removed, so the population remains constant.

\paragraph{Naive host.}
When a mutation appears, its average proportion at the end will remain at $1/N_{pop}$ (recall that the mutation is assumed to be neutral). In each realization, in the very long time, the mutation will be either lost of fixed. But the typical time for fixation will be of the order of $N_{pop}$ \citep{ewens2012mathematical}. So for a large $N_{pop}$, the mutations will stay a long time at a 0 or a small frequency. Thus, we will neglect the probability for a naive host to transmit more than one mutant. The probability to transmit a mutant will then be $N \mu G$, as in the mean field case. 

\paragraph{Immune host.}
In the case of an immune host, as in our model only one cluster is transmitted, we do not need to consider correlations between clusters. A cluster at generation $G$ was founded by one bacteria at generation $G-g$ (this bacteria may have been in a cluster at this point, but what matters is that at time $G$, all the bacteria in the cluster considered descend from this bacteria). Then, this cluster is fully mutant with probability $\mu (G-g)$, as $G-g$ is the number of replications between a bacteria in the inoculum and this founding bacteria. Besides, there is a probability $2\mu$ that a mutation occurred when this founding bacteria duplicated, $2^2\mu$ at next round... and $2^g \mu$ at the last round, and thus a probability $\mu(2+2^2+...+2^g)=\mu 2 (2^g-1)$ that a cluster is mixed. These are actually the same results as for the exponential case. 

\paragraph{Conclusion.}
To conclude, in the constant population size case, a transmission from a naive host will contain one mutant with probability $N \mu G$, as in the mean field case. A transmission from an immune host will involve a fully mutant cluster with probability $ \mu (G-g)$, and a mixed cluster with probability $2(N-1))$, exactly the same as for exponential growth. 

\subsubsection{Conclusion}

In all cases, for naive donor hosts, we assume that the $N$ transmitted bacteria are of types taken randomly and independently. In particular, when the initial bacteria are all of one type, if the average final proportion of mutants is $\rho$ (proportional to the mutation rate $\mu$, and thus very small), then the probability of transmitting one mutant bacteria among $N$ will be ${N\choose 1}\rho(1-\rho)^{N-1} \approx N \rho$, with very little chance of transmitting more than one mutant bacteria. 

For immune donor hosts, \emph{if the infection starts with a mixed inoculum}, we assume that all bacteria in a cluster transmitted to another host are of the same type. Conversely,  \emph{if the infection starts with an inoculum of bacteria that are all of the same type}, then other bacterial types are produced only by mutations. With $G$ the number of generations within the host and $N$ the bottleneck size and cluster size, if $G \gg N$, then most clusters will be of one bacterial type only, and the probability to transmit a fully mutant cluster will simply be the proportion of mutant bacteria, and a negligible amount of mixed clusters will be transmitted.  
If $G \ll N$, there are many more mixed clusters than clusters made of mutant bacteria only, and the proportion of mixed clusters is of the order of $2 N \mu$.

\subsection{Stochastic simulations}\label{simuSIresults}

Given our analytical arguments in section \ref{probaAuMoinsUnMut} (and in particular in section \ref{probaAuMoinsUnMutExp} which deals with the exponential growth assumed in our model), we expect the probability that a naive host initially infected by bacteria of a single type transmits at least one mutant to be very little modified by the fact that some mutations happen earlier than others due to the stochasticity of within-host growth. There could still be correlations between transmissions from the same host (if a mutation happened early within this host, then it transmits more mutants to all its contacts).

\begin{figure}[h t b]
\centering
\includegraphics[width=0.6\linewidth]{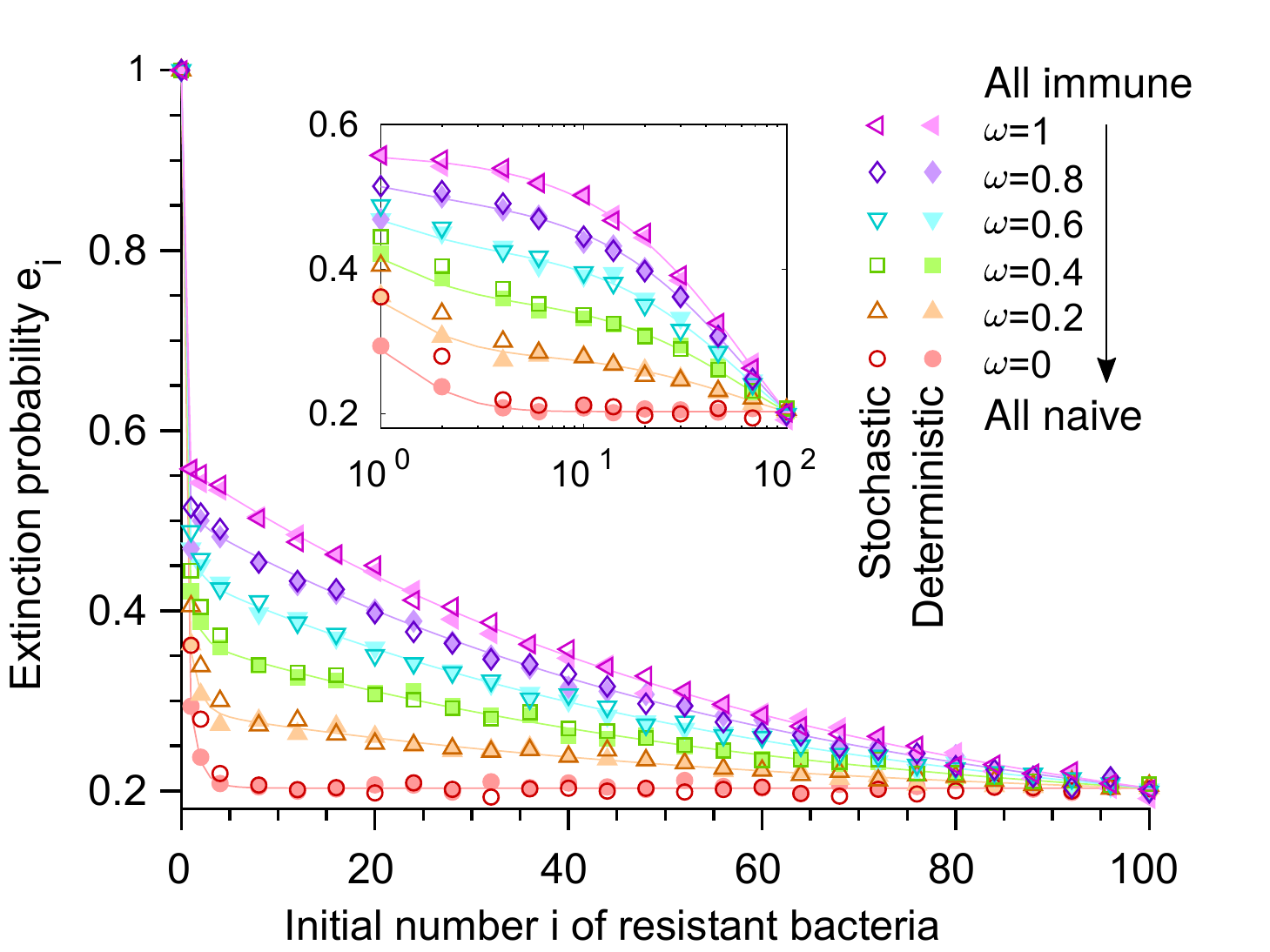}
\caption{\label{FigMutS}\textbf{Stochastic within-host growth without mutations or a fitness cost of resistance.} Extinction probabilities $e_i$ versus initial number $i$ of resistant bacteria in the first infected host, for different fractions $\omega$ of immune individuals in the host population.  Dark shades: results from simulations with stochastic mutation-free within-host growth. Light shades: results with deterministic mutation-free within-host growth (identical to Fig.~\ref{Fig1}). As in Fig.~\ref{Fig1}, $N=100$, $\lambda=2$, $q=0.55$, and resistance carries no cost. Solid lines: numerical resolution of Eq.~\eqref{thesystem}; symbols: simulation results (over $10^4$ realizations).}
\end{figure}

To address this question, we performed simulations with stochastic exponential within-host growth with no bacterial death (see Supporting Information, section~\ref{Simu_SI}). In this model without death, genotypes cannot fix inside a host (in contrast with models at fixed population size~\citep{Ewens79, Marrec18}). Nevertheless, stochastic exponential growth induces variability in the composition of bacterial populations growing from a given mixed inoculum, especially if it is small~\citep{Wienand15}. Fig.~\ref{FigMutS} compares results obtained with stochastic and deterministic within-host growth, without mutations. Overall, the impact of stochasticity is small, which validates our use of a deterministic model. Besides, the impact of stochasticity is strongest for naive hosts and small initial numbers $i$ of resistant bacteria. Again, non-treated immune hosts then transmit resistance very rarely anyway because of clustering. Conversely, for non-treated naive hosts with small $i$, exact population composition matters because it substantially affects the probability of transmitting resistance. Stochasticity can increase or decrease the resistant fraction, but decreasing it has more impact than increasing it, because beyond some small fraction, transmission of resistance is almost certain (see Eq.~\eqref{almost_eN}). Hence, stochasticity yields a slight increase of extinction probabilities in this case, as seen on Fig.~\ref{FigMutS}. 

\begin{figure}[h t b]
\centering
\begin{subfigure}[b]{0.49\linewidth}
  \centering
\includegraphics[width=\linewidth]{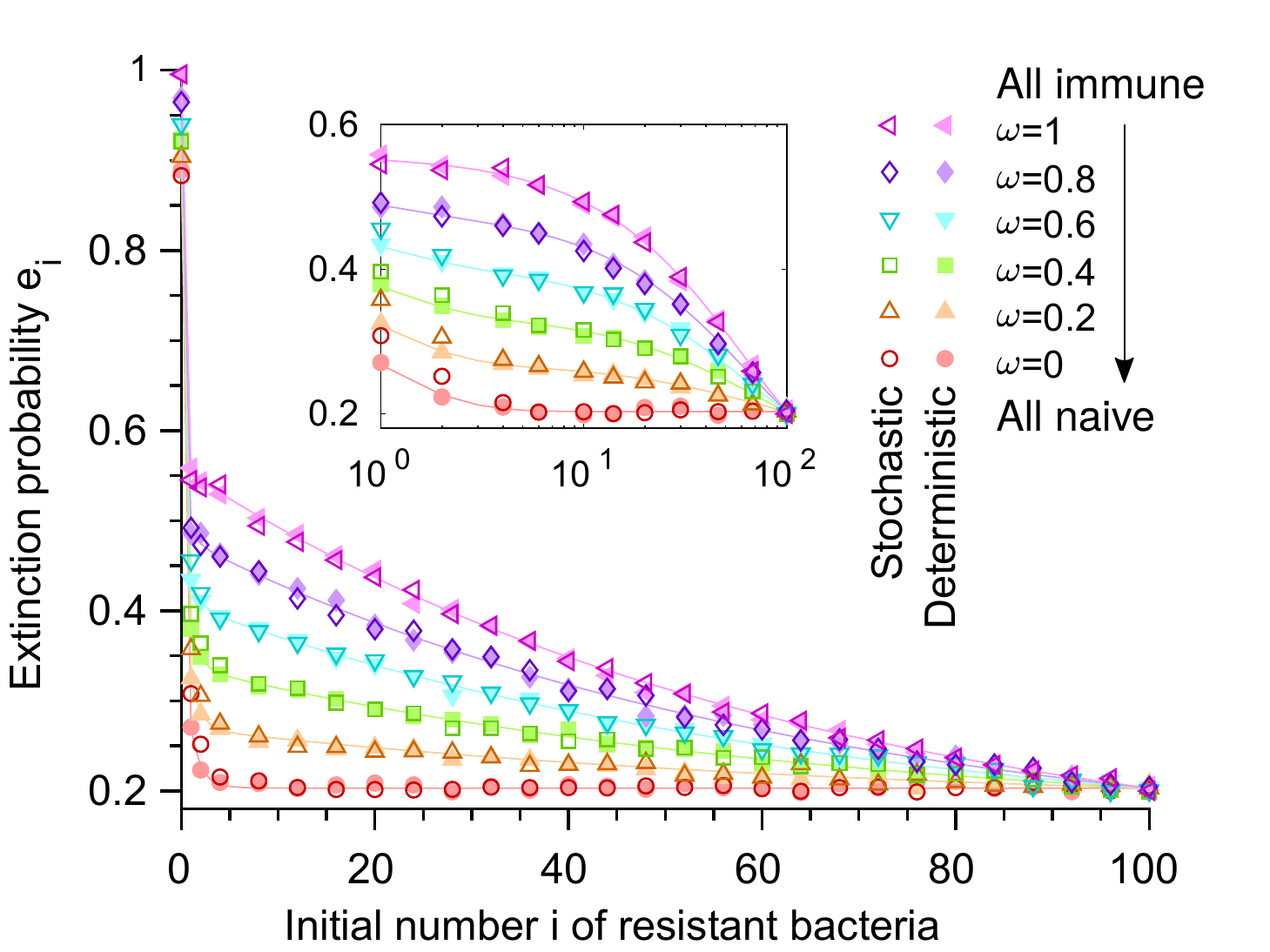}
\caption{}
\end{subfigure}
\begin{subfigure}[b]{0.49\linewidth}
  \centering
\includegraphics[width=\linewidth]{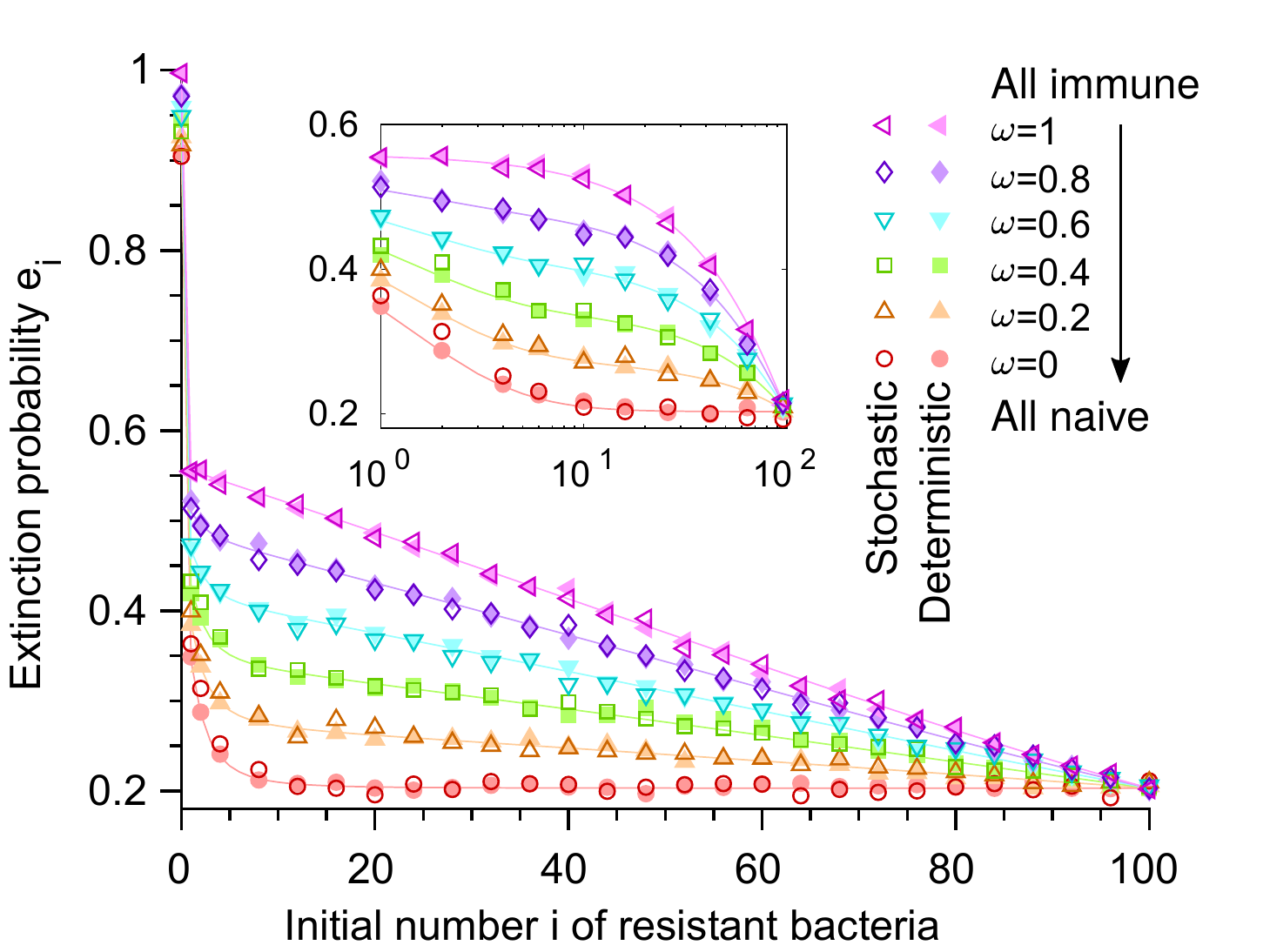}
\caption{}
\end{subfigure}
\caption{\label{FigMut2}\textbf{Stochastic within-host growth with  mutations and a fitness cost of resistance.} The extinction probability $e_i$ is shown as a function of the initial number $i$ of resistant bacteria in the first infected individual, for different values of the fraction $\omega$ of immune individuals in the host population. \textbf{A:} Case with mutations but no fitness cost. \textbf{B:} Case with mutations and a fitness cost $\delta =0.1$. In both panels, results from simulations with stochastic exponential within-host growth are shown in dark shades, and results with deterministic exponential within-host growth are shown in light shades for comparison (similarly to Fig.~\ref{FigMutS}). As in Fig.~\ref{Fig1}, we take a bottleneck size $N=100$, and each individual transmits bacteria to an average of $\lambda_\mathcal{N}=\lambda_\mathcal{I}=\lambda=2$ new hosts and is treated with probability $q_\mathcal{N}=q_\mathcal{I}=q=0.55$, irrespective of whether it is naive or immune. Here, we take mutation probabilities $\mu_1= 5\times 10^{-5}$ and $\mu_{-1}=0$, and incubation time corresponding to 10 generations of bacteria within the host. Solid lines correspond to numerical resolution of Eq.~\eqref{thesystem}, while symbols show results from numerical simulations of the branching process, computed over $10^4$ realizations. Main panel: linear scale; inset: semi-logarithmic scale.}
\end{figure}

Since mutations are rare events, it is also important to treat them in a stochastic way. As shown in Fig.~\ref{FigMut2}, the previous conclusions hold in this case, both with and without a fitness cost of resistance. In particular, the overall impact of stochasticity is small.

\clearpage

\begin{figure}[h t b]
\centering
\begin{subfigure}[b]{0.49\linewidth}
  \centering
\includegraphics[width=\linewidth]{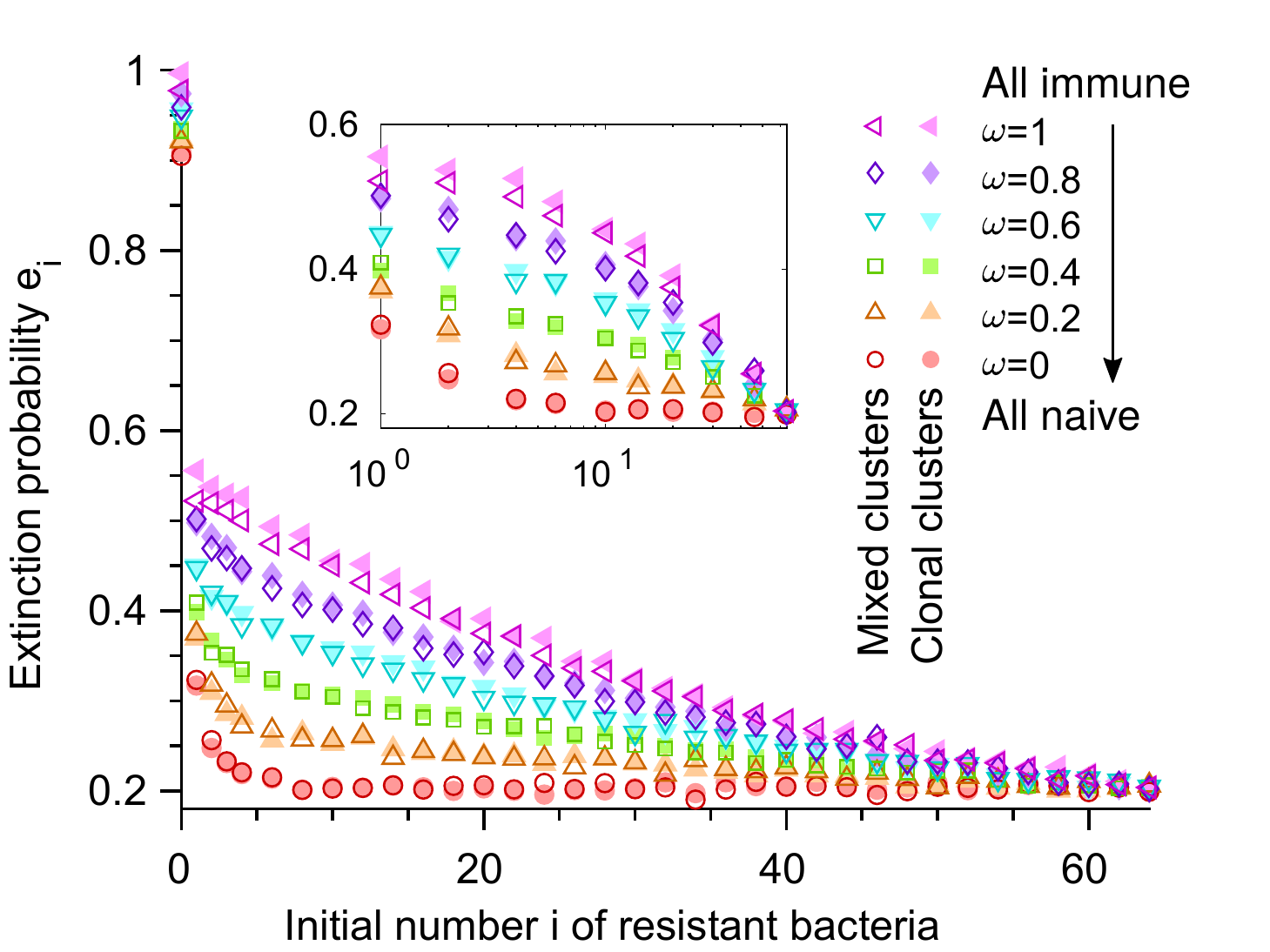}
\caption{}
\end{subfigure}
\begin{subfigure}[b]{0.49\linewidth}
  \centering
\includegraphics[width=\linewidth]{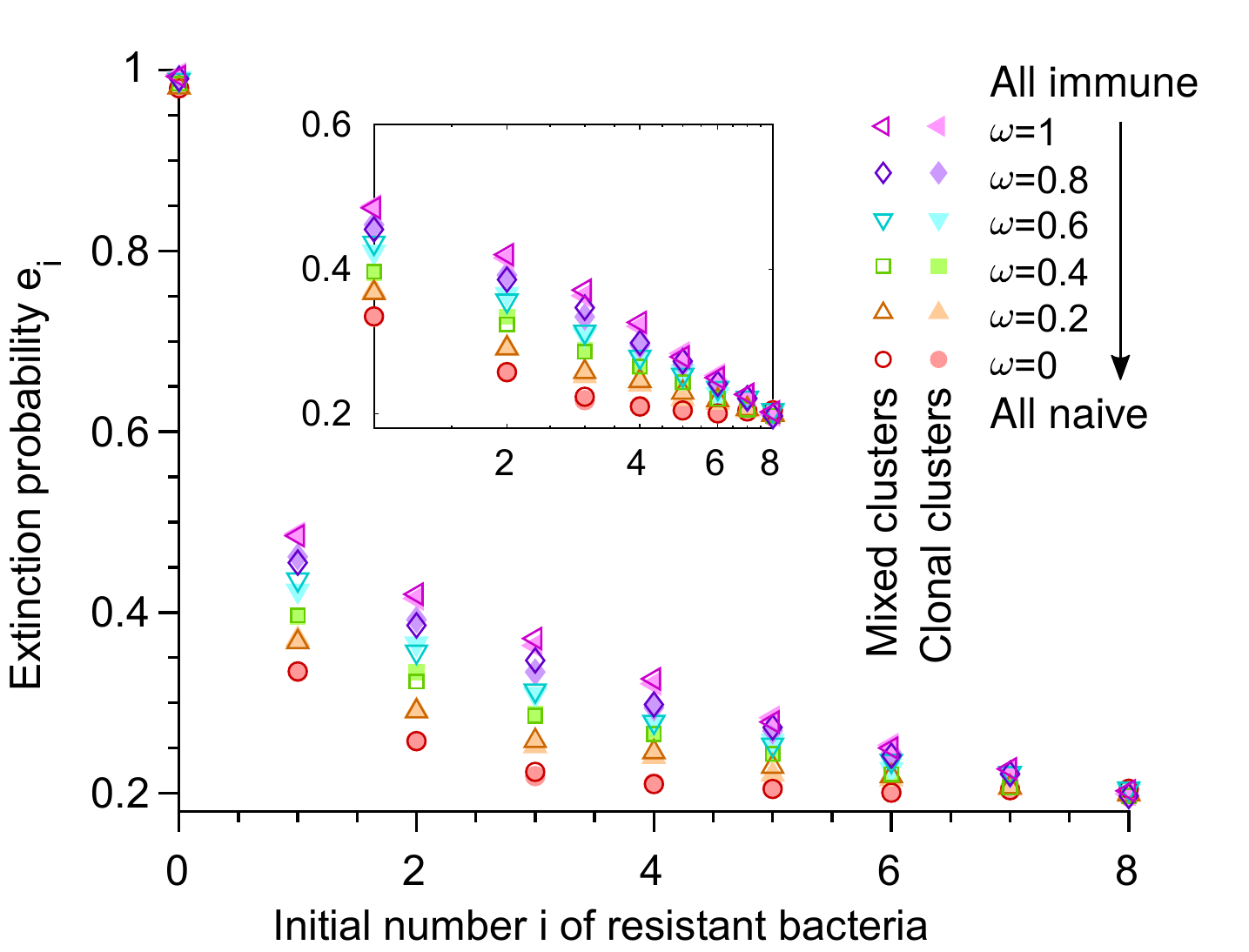}
\caption{}
\end{subfigure}
\caption{\label{FigMixedClust}\textbf{Impact of mixed clusters.} The extinction probability $e_i$ is shown as a function of the initial number $i$ of resistant bacteria in the first infected individual, for different values of the fraction $\omega$ of immune individuals in the host population. \textbf{A:} Cluster size $N=64$. \textbf{B:} Cluster size $N=8$. In both panels, results from simulations with stochastic exponential within-host growth including an explicit construction of clusters that allows for mixed clusters are shown in dark shades, and results with stochastic exponential within-host growth but with only clonal clusters (as in previous figures) are shown in light shades for comparison. As in Fig.~\ref{Fig1}, each individual transmits bacteria to an average of $\lambda_\mathcal{N}=\lambda_\mathcal{I}=\lambda=2$ new hosts and is treated with probability $q_\mathcal{N}=q_\mathcal{I}=q=0.55$, irrespective of whether it is naive or immune. Here, we take mutation probabilities $\mu_1= 5\times 10^{-5}$ and $\mu_{-1}=0$, and a total incubation time corresponding to $G=10$ generations of bacteria within the host. Symbols show results from numerical simulations of the branching process, computed over $10^4$ realizations. Main panel: linear scale; inset: semi-logarithmic scale.}
\end{figure}

We have also generally assumed that immune hosts transmit clusters comprising a single type of bacteria, except in our discussion of spread without any pre-existing resistance. As discussed above, immune hosts may also transmit mixed clusters. Fig.~\ref{FigMixedClust} shows that the influence of these mixed clusters is indeed often very small, especially for small cluster sizes (see Fig.~\ref{FigMixedClust}B). 

\newpage

\subsection{Impact of clustering on bacterial loss within a host}
\label{appendix_loss}

Throughout, we have considered no bacterial loss within a host, except in the section where we consider a population of constant size (section~\ref{cstpopsz}). Actually, some bacteria may be killed (for instance by bacteriophages, in which case bacteria in clusters, because they are very close in space, are very likely to be attacked together), or be lost in the feces (which would happen by clusters and not independently if bacteria are attached together). Let us look at the impact of including a small loss rate in the case where the proportion of resistant bacteria is initially small ($i \ll N$ resistant bacteria in the inoculum), and let us reason along the same lines as when deriving Eq.~\eqref{appx_eiI} of the main text. 

For a naive host, as we consider a small loss rate, it will be unlikely for the resistant bacteria to be lost before they replicate, and the more they are, the less likely it is that they will all be lost. Consequently, we expect very little influence of loss in a fully naive host population. 

In a fully immune population, let us start with a host infected with $i$ resistant bacteria. If this host is treated (probability $q_\mathcal{I}$), the infection becomes fully resistant, and leads to spread with probability $1-e_N^\mathcal{I}$. If this host is not treated, there is some probability that the resistant bacteria are lost. Because of clustering, this probability is higher than in the naive case; let us denote this probability $p_{i,loss}$. Then, if  $(1-q_\mathcal{I})\lambda_\mathcal{I}<1$, spread is possible only if resistant bacteria are not lost (probability $1-p_{i,loss}$), and when they are transmitted, they may lead to spread. This last term of spread is $1-\exp(-\lambda_{\mathcal{I}}(1-e_{each}))$, with $e_{each}$ the probability of extinction following each transmission. Here, we do not consider mutations, so there are no mixed clusters, and if a cluster of sensitive bacteria is transmitted, extinction is certain, because we assume  $(1-q_\mathcal{I})\lambda_\mathcal{I}<1$. If a cluster of resistant bacteria is transmitted, then it still leads to extinction with probability $e_N^\mathcal{I}$. The probability that the cluster transmitted is resistant is equal to the proportion of resistant bacteria, which is on average $i/N$ (unaffected by losses). Because in a proportion $p_{i,loss}$ of the cases, there is no resistant bacteria, then the proportion of resistant bacteria in the case that they are not lost will be $i/(N (1-p_{i,loss}))$. Then overall:   
\begin{equation}
1-e_i^\mathcal{I} \approx q_\mathcal{I} (1-e_N^\mathcal{I})+ (1-q_\mathcal{I})(1-p_{i,loss})(1-\exp(-\lambda_\mathcal{I}   (1-e_N^\mathcal{I})i /(N(1-p_{i,loss})))) 
\end{equation}
We are in the regime $i\ll N$, thus, assuming that $1-p_{i,loss}$ is not too small, $\lambda_\mathcal{I}   (1-e_N^\mathcal{I})i /(N(1-p_{i,loss})) \ll 1$, and consequently, 
\begin{equation}
1-e_i^\mathcal{I} \approx q_\mathcal{I} (1-e_N^\mathcal{I})+ (1-q_\mathcal{I})(1-p_{i,loss})(\lambda_\mathcal{I}   (1-e_N^\mathcal{I})i /(N(1-p_{i,loss}))))\approx  q_\mathcal{I} (1-e_N^\mathcal{I})+ (1-q_\mathcal{I}) \lambda_\mathcal{I}   (1-e_N^\mathcal{I})i/N\,.
\end{equation}
We thus recover Eq.~\eqref{appx_eiI} of the main text.  Therefore, provided that $p_{i,loss}$ is not too large (i.e. $1-p_{i,loss} \gg \lambda   (1-e_N^\mathcal{I})i / N $), including loss does not change our conclusions.

\section{Model with three bacterial types, including compensation}
\label{ThreeTypes}

Here, we extend our model to include the possibility of compensation of the fitness cost of resistance. This description involves three types of bacteria, sensitive ones (S) with division rate $f$ without antimicrobial, resistant ones (R) with division rate $f(1-\delta)$, and resistant-compensated ones (C) with division rate $f$. 

\subsection{Within-host growth}

Let us first write the within-host growth equations. For simplicity, we assume that there are no back-mutations. As in section~\ref{appendix_within-host_growth}, we present here the deterministic description of the within-host growth using ordinary differential equations, and assuming exponential growth. This yields the following system of linear differential equations:
\begin{equation}
\begin{cases}
\displaystyle
\dot{S}=f(1-\tilde{\mu}_1)S \\
\displaystyle
\dot{R}=f(1-\delta)(1-\tilde{\mu}_2)R+ f \, \tilde{\mu}_1 \, S \\
\displaystyle
\dot{C}=f \, C+ f(1-\delta) \, \tilde{\mu}_2 \, R\mbox{ },
\end{cases}
\label{deter_syst}
\end{equation}
where $S$, $R$ and $C$ are the numbers of sensitive (S), resistant (R) and resistant-compensated (C) microorganisms, respectively, and $f$ denotes the division rate of S bacteria in the absence of antibiotic, while dots denote time derivatives. Recall that R bacteria experience a fitness cost $\delta$, but C bacteria do not experience any cost. Also note that the mutation rates $\tilde{\mu}_i=\mu_i/\log(2)$ for $i=1,2$ are corrected to give agreement with the discrete model (see section~\ref{appendix_within-host_growth}).

Note that the population fractions $s=S/(S+R+C)$, $r=R/(S+R+C)$ and $c=C/(S+R+C)$ satisfy the following equations:
\begin{equation}
\begin{cases}
\displaystyle
\dot{s}=f(1-\tilde{\mu}_1)s-\overline{f}s \\
\displaystyle
\dot{r}=f(1-\delta)(1-\tilde{\mu}_2)r+ f \, \tilde{\mu}_1 \, s -\overline{f} \, r \\
\displaystyle
s+r+c=1 \mbox{ },
\end{cases}
\label{pop_frac}
\end{equation}
where $\overline{f}=f \, s + f(1-\delta) \, r +f \, c$ denotes the average fitness. Such equations are often taken as a starting point to describe large populations, and are known as replicator-mutator equations~\citep{Traulsen09}.

Being linear, the system in Eq.~\eqref{deter_syst} is straightforward to solve analytically:
\begin{equation}
\begin{pmatrix}
S \\ R \\ C
\end{pmatrix}
=
\begin{pmatrix}
0 & 0 & 1 \\
0 & 1 & \frac{\tilde{\mu}_1}{\delta-\tilde{\mu}_1+(1-\delta)\tilde{\mu}_2} \\
1 & -\frac{(1-\delta) \, \tilde{\mu}_2}{\delta+(1-\delta)\tilde{\mu}_2} & -\frac{(1-\delta) \, \tilde{\mu}_2}{\delta-\tilde{\mu}_1+(1-\delta)\tilde{\mu}_2}
\end{pmatrix}
\begin{pmatrix}
\alpha_1 \, e^{f \, t} \\ \alpha_2 \, e^{f(1-\delta) (1-\tilde{\mu}_2)t} \\ \alpha_3 \, e^{f (1-\tilde{\mu}_1)t}
\end{pmatrix} \mbox{ },
\label{gen_sol}
\end{equation}
where $\alpha_1$, $\alpha_2$ and $\alpha_3$ can be expressed from the initial conditions $S(0)$, $R(0)$ and $C(0)$. The fractions $s$, $r$ and $c$ can then be obtained from this solution, e.g. through $s=S/(S+R+C)$.

\subsection{Generating function}

Here, we present a full derivation of the generating function of the branching process in the case including three types of bacteria. For simplicity, we assume that there are no back-mutations and that all IgA-mediated clusters of bacteria are clonal.

Let $\wp_{ij}(\{n_{kl}\})$ be the probability that the first infected host, which is initially infected with $i$ R bacteria, $j$ C bacteria and $N-i-j$ S bacteria, transmits: \\
- 0 R bacteria, 0 C bacteria and $N$ S bacteria to $n_{00}$ hosts \\
- ... \\
- $k$ R bacteria, $l$ C bacteria and $N-k-l$ S bacteria to $n_{kl}$ hosts \\
- ... \\
Note that $0 \leq k,l \leq N$ and $0 \leq k+l \leq N$, where $N$ is the bottleneck size. Thus, $(N+1)(N+2)/2$ ordered pairs ($k,l$) are possible, and we will denote by $\mathcal{S}$ the set containing all of them. As in Eq.~\eqref{thesystem}, we distinguish four different cases, immune ($\mathcal{I}$) and naive ($\mathcal{N}$), treated ($t$) or non-treated ($nt$), so that:
\begin{align}
\wp_{ij}(\{n_{kl}\}) = & \, \underbrace{\omega \, q_\mathcal{I} \, \wp_{ij}^{\mathcal{I},t}(\{n_{kl}\})}_{Immune\mbox{ }and\mbox{ }treated}+\underbrace{(1-\omega)\, q_\mathcal{N} \, \wp_{ij}^{\mathcal{N},t}(\{n_{kl}\})}_{Naive\mbox{ }and\mbox{ }treated} \nonumber \\
& +\underbrace{\omega \, (1-q_\mathcal{I}) \, \wp_{ij}^{\mathcal{I},nt}(\{n_{kl}\})}_{Immune\mbox{ }and\mbox{ }untreated}+\underbrace{(1-\omega) \, (1-q_{\mathcal{N}}) \, \wp_{ij}^{\mathcal{N},nt}(\{n_{kl}\})}_{Naive\mbox{ }and\mbox{ }untreated}\, ,
\end{align}
where $\omega$ is the proportion of immune hosts in the population, and $q_{\mathcal{I}}$ and $q_{\mathcal{N}}$ are the probabilities of treatment for immune and naive hosts, respectively. 

The generating function of the branching process
\begin{equation} g_{ij}(\{z_{kl}\})=\sum_{\{n_{kl}/(k,l)\in\mathcal{S}\}}^{+\infty}\wp_{ij}(\{n_{kl}\})\prod_{(k,l)\in\mathcal{S}}z_{kl}^{n_{kl}}
\end{equation}
can thus be expressed as
\begin{align}
g_{ij}(\{z_{kl}\}) = & \, \underbrace{\omega \, q_\mathcal{I} \, g_{ij}^{\mathcal{I},t}(\{z_{kl}\})}_{Immune\mbox{ }and\mbox{ }treated}+\underbrace{(1-\omega)\, q_\mathcal{N} \, g_{ij}^{\mathcal{N},t}(\{z_{kl}\})}_{Naive\mbox{ }and\mbox{ }treated} \nonumber \\
& +\underbrace{\omega \, (1-q_\mathcal{I}) \, g_{ij}^{\mathcal{I},nt}(\{z_{kl}\})}_{Immune\mbox{ }and\mbox{ }untreated}+\underbrace{(1-\omega) \, (1-q_{\mathcal{N}}) \, g_{ij}^{\mathcal{N},nt}(\{z_{kl}\})}_{Naive\mbox{ }and\mbox{ }untreated} \, . \label{gen_fct}
\end{align}
Let us now present a derivation of the explicit form of the generating function in each of the four cases involved in this equation. Taken together, they will thus yield the complete form of the generating function, needed to compute the extinction probabilities.

\subsubsection{Immune and treated hosts}

\paragraph{Case $i=0$, $j=0$.}

In this case, we start with only sensitive bacteria, which are killed by the treatment. Thus, no bacteria can be transmitted. It follows that $\displaystyle \wp_{00}^{\mathcal{I},t}(\{n_{kl}\})=\prod_{(k,l)\in\mathcal{S}}\delta_{n_{kl},0}$ and that $g_{00}^{\mathcal{I},t}(\{z_{kl}\})=1$.

\paragraph{Case $i=0$, $j>0$.}

Since we assume that there are no back-mutations, the host will only harbor C bacteria and will transmit a cluster of $N$ C bacteria to each recipient host. Hence, $\displaystyle \wp_{0j}^{\mathcal{I},t}(\{n_{kl}\})=\frac{\lambda_\mathcal{I}^{n_{0N}}}{n_{0N}!}e^{-\lambda_\mathcal{I}}\prod_{(k,l) \in \mathcal{S} \backslash \{(0,N)\}}\delta_{n_{kl},0}$ and $\displaystyle g_{0j}^{\mathcal{I},t}(\{z_{kl}\})=\exp\left(-\lambda_\mathcal{I}(1-z_{0N})\right)$.

\paragraph{Case $i>0$, $j\geq0$.}

In this case, we have two types of bacteria (R and C). Even if initially $j=0$, C bacteria may appear by mutation, and their number may be nonzero after the incubation time $\tau$. Because there are no back-mutations, S bacteria cannot reappear after being killed by the antimicrobial. Since we neglect mixed clusters, recipient hosts can be contaminated only by clonal clusters of $N$ R bacteria or $N$ C bacteria. The probability that $n_{N0}$ clusters of R bacteria only are transmitted, given that $n$ clusters are transmitted, reads
\begin{equation}
p(n_{N0}|n)={n\choose{n_{N0}}}(r_{i,j}^{i+j})^{n_{N0}}(1-r_{i,j}^{i+j})^{n-n_{N0}} \, . 
\end{equation}
where $r_{i,j}^{i+j}$ (resp. $c_{i,j}^{i+j}=1-r_{i,j}^{i+j}$) represents the fraction of R bacteria (resp. C bacteria) at the end of the incubation period, starting from an initial total number of bacteria $i+j$, an initial number of R bacteria $i$  and an initial number $j$ of C bacteria. 
Using the law of total probability, we then obtain the probability that $n_{N0}$ clusters of R bacteria only are transmitted: 
\begin{equation}
p(n_{N0})  = \sum_{n=n_{N0}}^{+\infty} p(n_{N0}|n)p(n) = \frac{(\lambda_\mathcal{I}\, r_{i,j}^{i+j})^{n_{N0}}}{n_{N0}!}\exp\left(-\lambda_\mathcal{I}\, r_{i,j}^{i+j}\right) \, , \label{LTP}
\end{equation}
where we have used $p(n)=\lambda_\mathcal{I}^n e^{-\lambda_\mathcal{I}}/n!$ since transmission is assumed to be Poissonian. Similarly, we obtain the probability that $n_{0N}$ clusters of C bacteria only are transmitted:
\begin{equation}
p(n_{0N})=\frac{(\lambda_\mathcal{I} (1-r_{i,j}^{i+j}))^{n_{0N}}}{n_{0N}!}\exp\left(-\lambda_\mathcal{I}(1-r_{i,j}^{i+j})\right) \, .
\end{equation}
We can then write $\displaystyle \wp_{ij}^{\mathcal{I},t}(\{n_{kl}\})=p(n_{N0})p(n_{0N})\prod_{(k,l) \in \mathcal{S} \backslash \{(0,N),(N,0)\}} \delta_{n_{kl},0}$, and thus \begin{equation} g_{ij}^{\mathcal{I},t}(\{z_{kl}\})=\exp\left(-\lambda_\mathcal{I}\, r_{i,j}^{i+j}(1-z_{N0})\right)\exp\left(-\lambda_\mathcal{I}(1-r_{i,j}^{i+j})(1-z_{0N})\right)\, .
\end{equation}

\subsubsection{Naive and treated hosts}

\paragraph{Case $i=0$, $j=0$.}

This case is identical to that of immune and treated hosts (see above) and $g_{00}^{\mathcal{N},t}(\{z_{kl}\})=1$.

\paragraph{Case $i=0$, $j>0$.}

This case is identical to that of immune and treated hosts (see above), except that the host transmits to an average of $\lambda_\mathcal{N}$ new hosts (instead of $\lambda_\mathcal{I}$). Hence $\displaystyle g_{0j}^{\mathcal{N},t}(\{z_{kl}\})=\exp\left(-\lambda_\mathcal{N}(1-z_{0N})\right)$. 

\paragraph{Case $i>0$, $j\geq0$.}

As explained for immune and treated hosts, in this case, we have two types of bacteria (R and C). Even if initially $j=0$, C bacteria may appear by mutation, and their number may be nonzero after the incubation time $\tau$. Because there are no back-mutations, S bacteria cannot reappear after being killed by the antimicrobial.

Let us denote by $n$ the total number of new hosts that will be infected by the host considered. In principle, we should draw $n$ samples of $N$ bacteria without replacement out of the $N^2$ bacteria assumed to be present at the end of the incubation time. The composition of each sample will potentially impact the others. However, this effect will be negligible if $N \gg n$, in which case we can consider for simplicity that each of the $n$ samples is drawn with replacement from the set of $N^2$ bacteria. Within this approximation, the probability that there are $k$ R bacteria in a packet of $N$ bacteria follows a binomial law:
\begin{equation}
B_{i,j}(k)={N \choose i}(r_{i,j}^{i+j})^k(1-r_{i,j}^{i+j})^{N-k}\, .
\end{equation}
where we have used $c_{i,j}^{i+j}=1-r_{i,j}^{i+j}$. The probability that $n_{k \, N-k}$ clusters including $k$ R bacteria and $N-k$ C bacteria are transmitted, given that $n$ clusters are transmitted reads:
\begin{equation}
p(n_{k \, N-k}|n) = {n \choose n_{k \, N-k}}(B_{i,j}(k))^{n_{k \, N-k}}(1-B_{i,j}(k))^{n-n_{k \, N-k}} \, .
\end{equation}
Using the law of total probability (see Eq.~\eqref{LTP}), we then obtain the probability that $n_{k \, N-k}$ clusters including $k$ R bacteria and $N-k$ C bacteria are transmitted:
\begin{equation}
p(n_{k \, N-k})= \frac{(\lambda_\mathcal{N}B_{i,j}(k))^{n_{k \, N-k}}}{n_{k \, N-k}!}\exp\left(-\lambda_\mathcal{N}B_{i,j}(k)\right) \, .
\end{equation}
Since $\displaystyle \wp_{ij}(\{n_{kl}\})=\prod_{k=0}^N p(n_{k \, N-k})$, we obtain:
\begin{align}
g_{ij}^{\mathcal{N},t}(\{(z_{kl})\})& =\sum_{n_{kl}}\left(\prod_{k=0}^N p(n_{k \, N-k})\right)\left(\prod_{(k,l)\in \mathcal{S} \backslash \{(k,l)/k+l=N\}}\delta_{n_{kl},0}\right)\left( \prod_{(k,l)\in\mathcal{S}}z_{kl}^{n_{kl}}\right) \nonumber\\ 
& = \prod_{k=0}^N \left( \sum_{n_{k \, N-k}} p(n_{k \, N-k}) z_{k \, N-k}^{n_{k \, N-k}} \right) \nonumber\\ 
& = \prod_{k=0}^N \left( \sum_{n_{k \, N-k}} \frac{(\lambda_\mathcal{N}B_{i,j}(k))^{n_{k \, N-k}}}{n_{k \, N-k}!}\exp\left(-\lambda_\mathcal{N}B_{i,j}(k)\right) z_{k \, N-k}^{n_{k \, N-k}} \right) \nonumber\\ 
& =\exp \left(-\lambda_\mathcal{N} \sum_{k=0}^N B_{i,j}(k)(1-z_{k \, N-k})\right) \, .
\end{align}

\subsubsection{Immune and non-treated hosts}

Here, transmission involves clusters of $N$ identical bacteria of type S, R or C. Let us express the probability that $n_{N0}$ clusters of R bacteria, $n_{0N}$ clusters of C bacteria and $n_{00}=n-n_{N0}-n_{0N}$ of S bacteria are transmitted, given that $n$ clusters are transmitted:
\begin{equation}
p(n_{N0},n_{0N}|n)=\frac{n!}{n_{N0}!n_{0N}!(n-n_{N0}-n_{0N})!}(r_{i,j}^N)^{n_{N0}}(c_{i,j}^N)^{n_{0N}}(1-r_{i,j}^N-c_{i,j}^N)^{n-n_{N0}-n_{0N}} \, ,
\end{equation}
which is a trinomial distribution. Then it follows:
\begin{align}
p(n_{N0}|n)&=\sum_{n_{0N}=0}^{n-n_{N0}}p(n_{N0},n_{0N}|n)=\frac{n!}{(n-n_{N0})!n_{N0}!}(r_{i,j}^N)^{n_{N0}}(1-r_{i,j}^N)^{n-n_{N0}} \, .
\end{align}
Using the law of total probability (see Eq. \eqref{LTP}) yields
\begin{equation}
p(n_{N0})=\frac{\left(\lambda_\mathcal{I}\, r_{i,j}^N\right)^{n_{N0}}}{n_{N0}!}\exp\left(-\lambda_\mathcal{I}\, r_{i,j}^N\right) \, .
\end{equation}
Similarly, 
\begin{align} 
p(n_{0N})&=\frac{1}{n_{0N}!}\left(\lambda_\mathcal{I}\, c_{i,j}^N\right)^{n_{0N}}\exp\left(-\lambda_\mathcal{I}\, c_{i,j}^N\right)\,,\\
 p(n_{00})&=\frac{1}{n_{00}!}\left(\lambda_\mathcal{I}(1-r_{i,j}^N-c_{i,j}^N)\right)^{n_{00}}\exp\left(-\lambda_\mathcal{I}(1-r_{i,j}^N-c_{i,j}^N)\right)\,.
\end{align}
Since $\displaystyle \wp_{ij}^{\mathcal{I},nt}(\{n_{kl}\})=p(n_{00})p(n_{N0})p(n_{0N})\prod_{(k,l)\in\mathcal{S}\backslash \{(0,0),(N,0),(0,N)\}} \delta_{n_{kl},0}\,$, we finally obtain
\begin{align}
g_{ij}^{\mathcal{I},nt}(\{z_{kl}\})&= \exp\left(-\lambda_\mathcal{I}(1-r_{i,j}^N-c_{i,j}^N)(1-z_{00})\right)\nonumber\\ 
&\times \exp\left(-\lambda_\mathcal{I}\, r_{i,j}^N(1-z_{N0})\right) \times \exp\left(-\lambda_\mathcal{I}\, c_{i,j}^N(1-z_{0N})\right) \, . 
\end{align}

\subsubsection{Naive and non-treated hosts}

Here, transmission involves random assortments of N bacteria that may potentially contain all three different types of bacteria. The probability of having $k$ R bacteria and $l$ C bacteria in an assortment reads
\begin{equation}
T_{i,j}(k,l) = \frac{N!}{k!\,l!\,(N-k-l)!}(r_{i,j}^N)^k(c_{i,j}^N)^l(1-r_{i,j}^N-c_{i,j}^N)^{N-k-l} \, .
\end{equation}
Then we can express the probability of transmitting $n_{kl}$ packets with $k$ R bacteria and $l$ C bacteria, given that $n$ packets are transmitted:
\begin{equation}
p(n_{kl}|n)={n \choose n_{kl}}(T_{i,j}(k,l))^{n_{kl}}(1-T_{i,j}(k,l))^{n-n_{kl}} \, .
\end{equation}
Using the law of total probability (see Eq.~\eqref{LTP}) yields
\begin{equation}
p(n_{kl})=\frac{(\lambda_\mathcal{N}T_{i,j}(k,l))^{n_{kl}}}{n_{kl}!}\exp\left(-\lambda_\mathcal{N}T_{i,j}(k,l)\right) \, .
\end{equation}
Since $\displaystyle \wp_{ij}^{\mathcal{N},nt}(\{n_{kl}\})=\prod_{(k,l)\in \mathcal{S}}p(n_{kl})$, 
we finally obtain
\begin{equation}
g_{ij}^{\mathcal{N},nt}(\{z_{kl}\})=\exp\left(-\lambda_\mathcal{N}\sum_{(k,l)\in \mathcal{S}}T_{i,j}(k,l)(1-z_{kl})\right)\,.
\end{equation}

\subsubsection{Conclusion}

Combining the above results for the generating function in the various cases studied, we can explicitly rewrite Eq. \eqref{gen_fct}. This allows us to compute the extinction probabilities of epidemics $e_{ij}$, which are the fixed points of the generating function $g_{ij}(\{z_{kl}\})$, where the first host is initially infected by $i$ R bacteria, $j$ C bacteria and $N-i-j$ S bacteria.

\subsection{Results with three bacterial types}

\begin{figure}[h t b]
\centering
\includegraphics[width=0.6\linewidth]{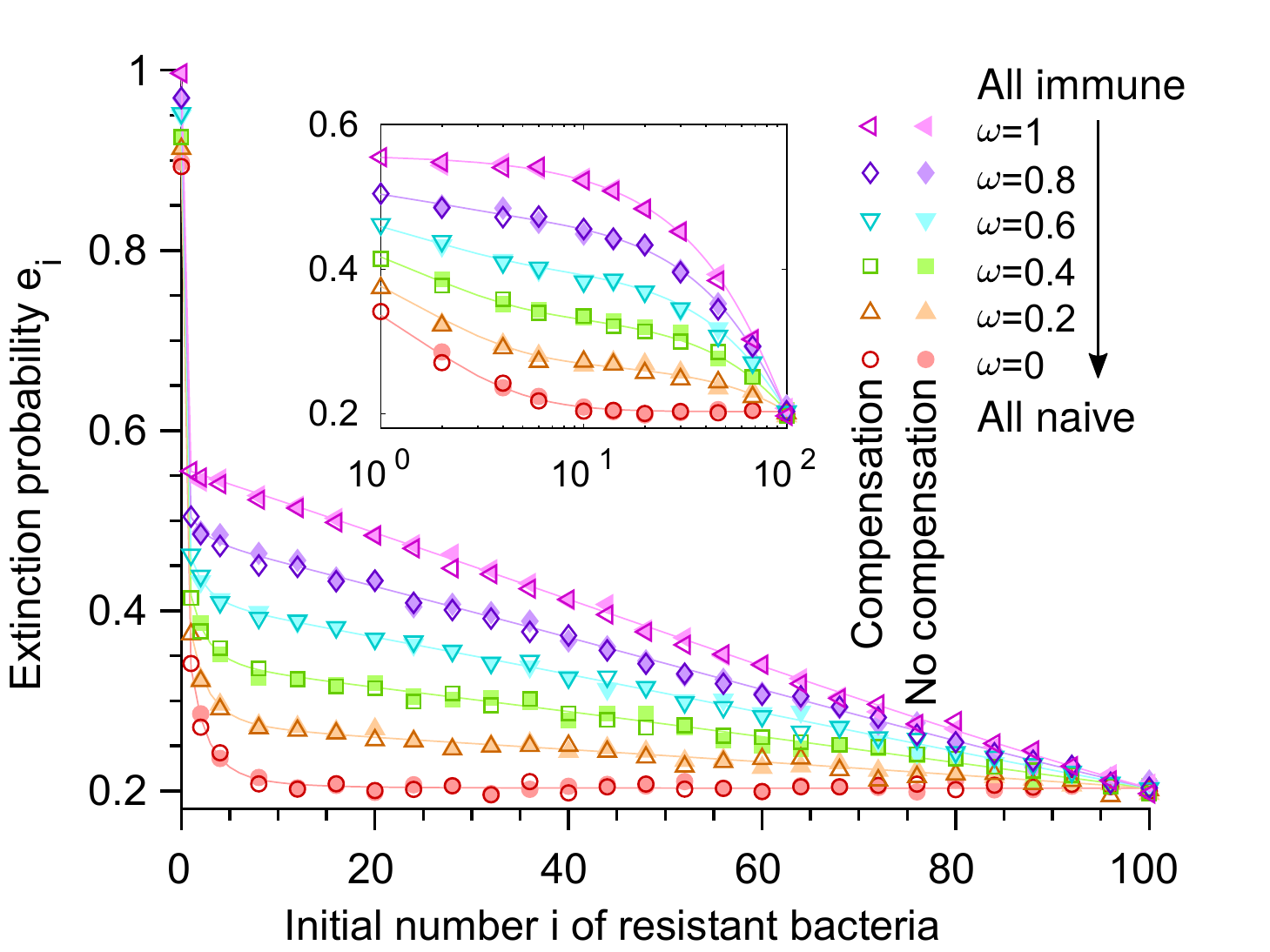}
\caption{\label{FigCostS} \textbf{Impact of compensation.} Extinction probabilities $e_i$ versus initial number $i$ of resistant bacteria in the first infected host, for different fractions $\omega$ of immune individuals in the host population. Dark shades: results with a fitness cost $\delta =0.1$ and compensation with $\mu_2=7\times 10^{-3}$. Light shades: results with $\delta =0.1$ but no compensation (same as dark-shaded results in Fig.~\ref{FusionFig}B). Within-host evolution is deterministic, $\mu_1=7 \times 10^{-5}$ and $\mu_{-1}=0$, and incubation time is $G=10$ generations. As in Fig.~\ref{Fig1}, $N=100$, $\lambda=2$ and $q=0.55$. Solid lines: numerical resolution of Eq.~\eqref{thesystem}; symbols: simulation results (over $10^4$ realizations).}
\end{figure}

\begin{figure}[h t b]
\centering
\begin{subfigure}[b]{0.49\linewidth}
  \centering
\includegraphics[width=\linewidth]{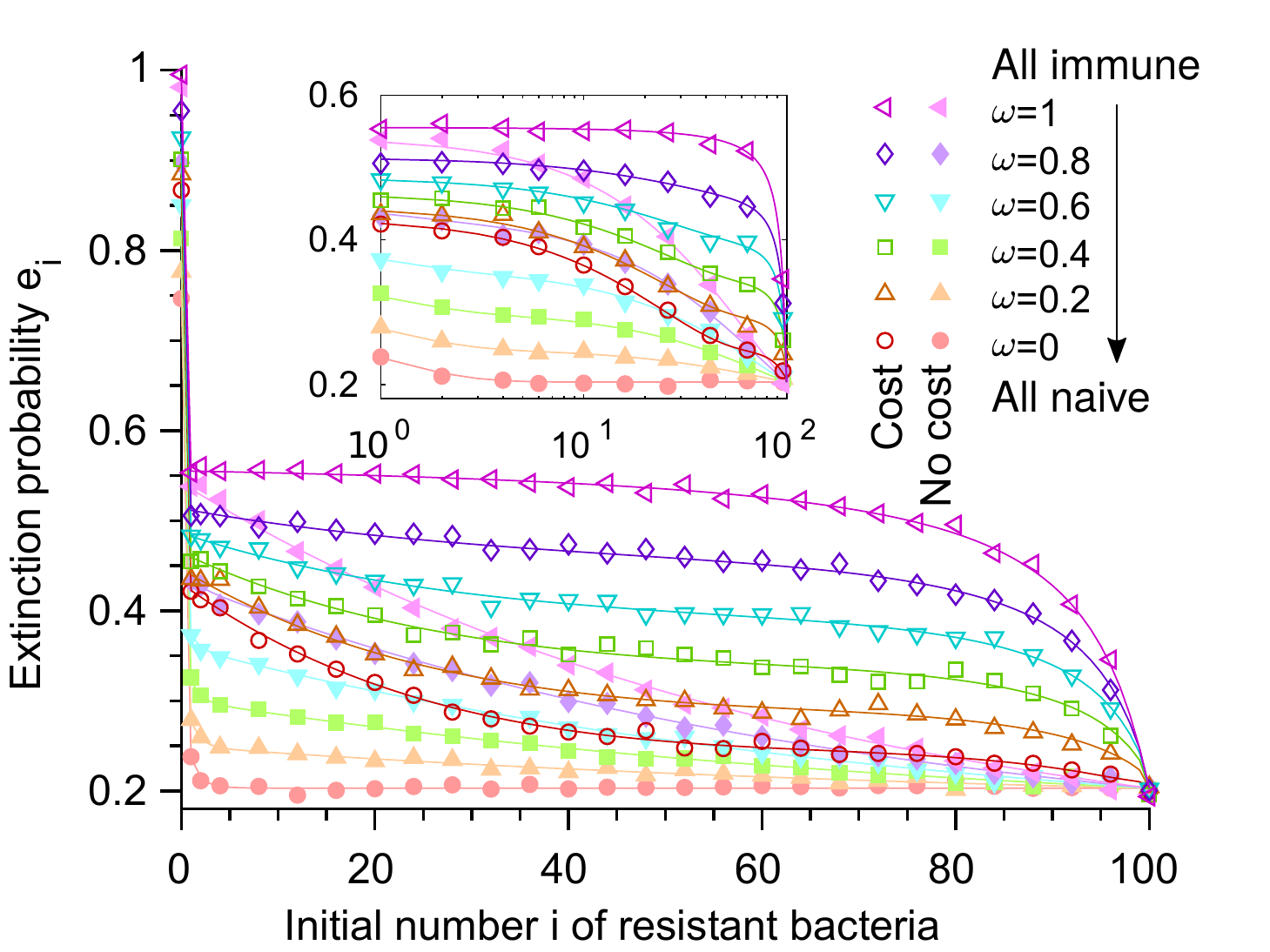}
\caption{}
\end{subfigure}
\begin{subfigure}[b]{0.49\linewidth}
  \centering
\includegraphics[width=\linewidth]{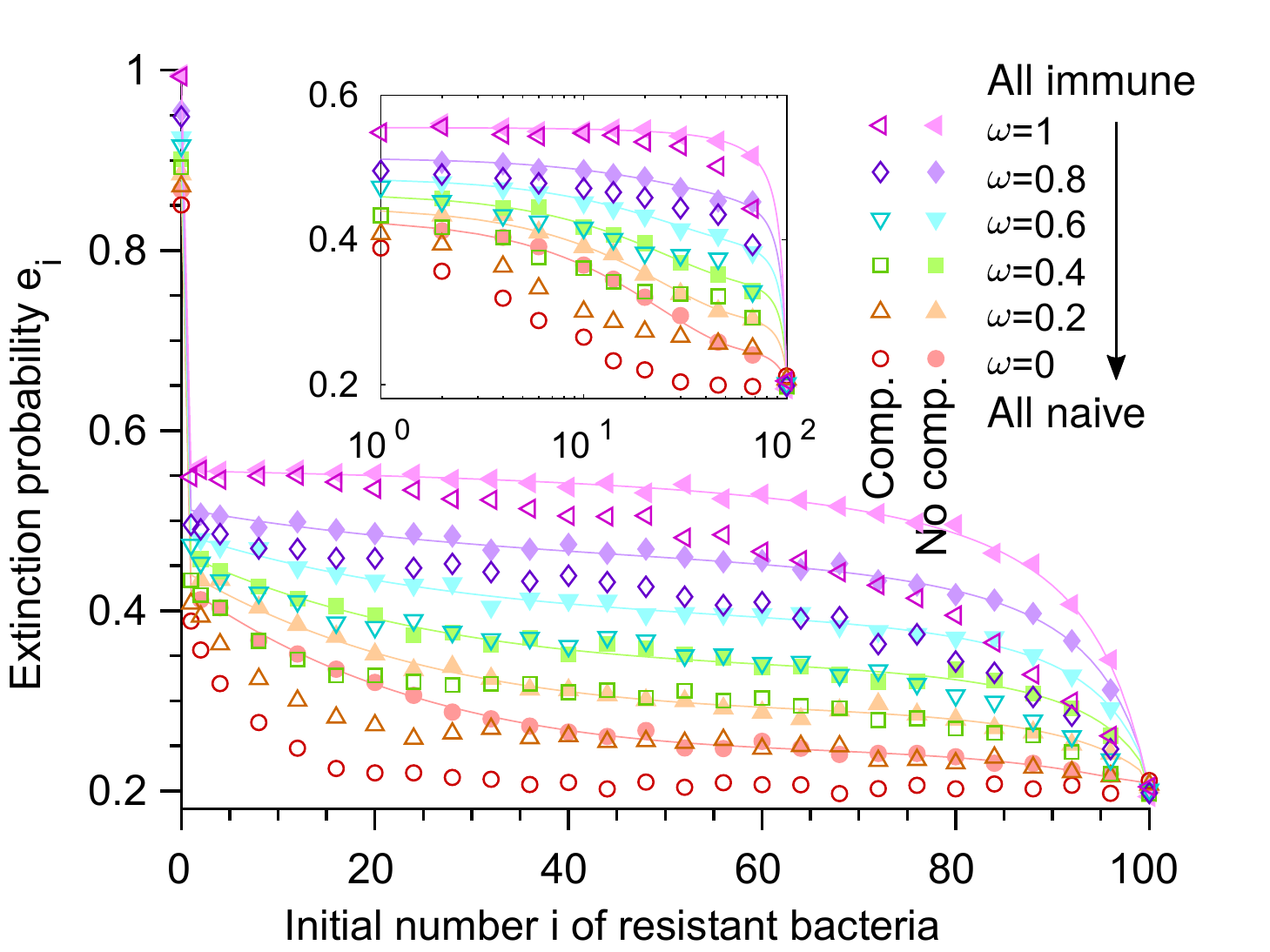}
\caption{}
\end{subfigure}
\caption{\label{FigCost2}\textbf{Impact of a fitness cost of resistance and of compensation for long incubation times.} The extinction probability $e_i$ is shown as a function of the initial number $i$ of resistant bacteria in the first infected individual, for different values of the fraction $\omega$ of immune individuals in the host population. \textbf{A:} Results with a fitness cost $\delta =0.1$ of resistance are shown in dark shades. Results without fitness cost (identical to the dark-shaded results in Fig.~\ref{FusionFig}A) are shown in light shades for comparison. \textbf{B:} Results with a fitness cost $\delta =0.1$ of resistance and with a possible compensatory mutation with mutation probabilities $\mu_2=7\times 10^{-3}$ and $\mu_{-2}=0$ are shown in dark shades. Results with the same fitness cost but without compensation (identical to the dark-shaded results in panel A) are shown in light shades for comparison. In both panels, within-host evolution is deterministic, mutation probabilities are $\mu_1=7 \times 10^{-5}$ and $\mu_{-1}=0$, and incubation time corresponds to 50 generations of bacteria within the host (instead of 10 generations in Figs.~\ref{FusionFig} and~\ref{FigCostS}). In addition, as in Fig.~\ref{Fig1}, we take a bottleneck size $N=100$, and each individual transmits bacteria to an average of $\lambda_\mathcal{N}=\lambda_\mathcal{I}=\lambda=2$ new hosts and is treated with probability $q_\mathcal{N}=q_\mathcal{I}=q=0.55$, irrespective of whether it is naive or immune. Solid lines correspond to numerical resolution of Eq.~\eqref{thesystem} (not shown with compensation), while symbols show results from numerical simulations of the branching process, computed over $10^4$ realizations. Main panel: linear scale; inset: semi-logarithmic scale.}
\end{figure}

Because multiple mutations can compensate for the initial cost of resistance~\citep{Levin00,Paulander07,Hughes15}, the probability $\mu_2$ of compensatory mutations often satisfies $\mu_2\gg\mu_1$~\citep{Levin00,Poon05}. In Fig.~\ref{FigCostS}, compensation has a negligible impact. Indeed, the proportion of compensated bacteria then remains small. However, for longer incubation times, compensation decreases extinction probabilities, which become intermediate between those with a cost and those without a cost (see Fig.~\ref{FigCost2}). Overall, compensation does not have a major impact on the initial steps of the propagation studied here.

\clearpage

\newpage

\section{Detailed simulation methods}
\label{Simu_SI}

Our code is freely available at \texttt{https://doi.org/10.5281/zenodo.2592323}.

Our simulations were coded in Matlab. Here, we present in detail our simulation scheme, which is outlined in the Methods section of the main text. We perform stochastic simulations of the branching process modeling the propagation of the bacterial strain considered. The within-host growth can be treated either deterministically or stochastically, and in the latter case, it can either allow mixed clusters or not (see below). We consider ``generations'' of hosts, the first one corresponding to the first infected host, the second one to the hosts directly infected by this first host, etc. Throughout this section, as in our simulations, we assume that there are no back-mutations.

\subsection{Initialization (first host)} 

We start with a first generation where there is a single host infected by an inoculum containing $i$ R bacteria, $j$ C bacteria and $N-i-j$ S bacteria. 

\subsection{Loop over successive generations of hosts}

For each host in the generation considered, we first randomly choose whether it is immune or naive, according to the probability $\omega$ of being immune. 

\subsubsection{Naive hosts}

For each naive host in the generation considered, we randomly choose whether it is treated or not, according to the probability $q_\mathcal{N}$ of being treated.

		\begin{itemize}
			\item If the naive host is treated:
			\begin{itemize}
				\item \textbf{Treatment:} All sensitive (S) bacteria are eliminated from the inoculum.
				\item If the inoculum contained at least one resistant (R) or compensated (C) bacterium:
				\begin{enumerate}
					\item \textbf{Within-host growth:} In the absence of compensation, since all S bacteria were eliminated and since we consider no back-mutation, only R bacteria will exist within the host. If compensation is accounted for, the fractions of R and C bacteria at the end of the incubation period are computed using either a deterministic or a stochastic approach (see below). Then, we compute the numbers of R and C bacteria by randomly sampling them in a binomial distribution with parameters $N^2$ (number of bacteria after the within-host growth) and $r$ (fraction of R bacteria after the within-host growth).
					\item \textbf{Transmission to the next generation of hosts:} The number of hosts infected by the present host is computed by a random draw from a Poisson distribution with average $\lambda_\mathcal{N}$. Each new infected host receives a set of $N$ microorganisms, randomly drawn without replacement from the $N^2$ bacteria infecting the transmitting host at the end of the incubation time.
				\end{enumerate}
			\item If the inoculum only contained S bacteria, no transmission occurs, since the treatment eliminated all bacteria.
			\end{itemize}
		\item If the naive host is not treated:
		\begin{enumerate}
			\item \textbf{Within-host growth:} The proportions of S and R bacteria (and C if compensation is accounted for) at the end of the incubation period are computed using either a deterministic or a stochastic approach (see below). Then, in the absence of compensation, we compute the numbers of S and R bacteria by randomly sampling them in a binomial distribution with parameters $N^2$ and $s$. If compensation is accounted for, we instead use a trinomial distribution with parameters $N^2$, $s$ and $r$.
			\item \textbf{Transmission to the next generation of hosts} is performed as in the naive-treated case presented above.
		\end{enumerate}
		\end{itemize}
		
\subsubsection{Immune hosts}

For each immune host in the generation considered, we randomly choose whether it is treated or not, according to the probability $q_\mathcal{I}$ of being treated.

		\begin{itemize}
			\item If the immune host is treated:
			\begin{itemize}
			    \item \textbf{Treatment:} All sensitive (S) bacteria are eliminated from the inoculum.
				\item If the inoculum contained at least one R mutant or one C bacterium:
				\begin{enumerate}
					\item \textbf{Within-host growth:} Again, in the absence of compensation, only R bacteria will exist. With compensation, the fractions of R and C bacteria at the end of the incubation period are computed as in the naive-treated case. If only clonal clusters are considered, $N$ clonal clusters of $N$ bacteria are constructed by randomly drawing the number of R and C bacteria seeding clusters in a binomial distribution with parameters $N$ and $r$. If mixed clusters are considered, they are explicitly constructed at the end of the within-host growth (see below).
					\item \textbf{Transmission to the next generation of hosts:} The number of hosts infected by the present host is computed by a random draw from a Poisson distribution with average $\lambda_\mathcal{I}$. Each newly infected host receives one random cluster among the $N$ clusters formed at the end of the within-host growth.
				\end{enumerate}
				\item If the inoculum only contained S bacteria, no transmission occurs as the treatment eliminated all bacteria.
			\end{itemize}
			\item If the immune host is not treated:
			\begin{enumerate}
				\item \textbf{Within-host growth:} The proportions of S and R bacteria (and C if compensation is accounted for) at the end of the incubation period are computed as in the naive-non treated case. If only clonal clusters are considered, in the absence of compensation, we compute the numbers of S and R bacteria seeding clusters by randomly sampling them in a binomial distribution with parameters $N$ and $s$. If only clonal clusters are considered, but compensation is accounted for, we instead use a trinomial distribution with parameters $N$, $s$ and $r$. If mixed clusters are considered, they are explicitly constructed at the end of the within-host growth (see below).
				\item \textbf{Transmission to the next generation of hosts} is performed as in the immune-treated case.
			\end{enumerate}
		\end{itemize}

\subsection{End of the simulation} 

The simulation is stopped when one of the following conditions is met:
	\begin{itemize}
		\item Extinction, i.e. no more infected hosts at a given generation.
		\item The number of infected hosts at the current generation of hosts is larger than a threshold ($100$ in practice).
		\item The number of generations of infected hosts is larger than a threshold ($100$ in practice).
	\end{itemize}
We approximate the last two cases as meaning that no extinction will occur. We have checked that increasing the threshold values above 100 does not significantly affect our results, which confirms that this approximation is valid.

\subsection{Within-host growth}

\subsubsection{Deterministic model}

If the within-host growth is treated deterministically, in the absence of mutations and fitness cost, the proportion of resistant bacteria just remains the same upon growth. 

With mutations and/or fitness cost, we compute the fractions of sensitive and resistant bacteria (and compensated ones, if compensation is accounted for) at a time $\tau=G
\log(2)$ (see Eq.~\eqref{corrGtau} with $f=1$) equivalent to the total number of generations $G$ of the incubation period, using the solution of our ODE model (see section~\ref{appendix_within-host_growth}, especially Eqs.~\eqref{ode1} and~\eqref{eqRdebut} for the case without compensation and section~\ref{ThreeTypes}, especially Eq.~\eqref{deter_syst} for the case with compensation). Note that even though the within-host growth is treated deterministically, the sampling of transmitted bacteria and the branching process are stochastic.

\subsubsection{Stochastic model}

If the within-host growth is treated stochastically, we implement an exact Gillespie simulation scheme~\citep{Gillespie76,Marrec18} for a fixed number $N_{div}=N(2^G-1)$ of single bacterial divisions, corresponding to the number of divisions that take place during the $G$ generations of the incubation period. In our Gillespie algorithm, each bacterium can divide randomly with a rate equal to its fitness, and each daughter cell can then mutate randomly with a given probability ($\mu_1$ for $\mbox{S}\rightarrow\mbox{R}$ and $\mu_2$ for $\mbox{R}\rightarrow\mbox{C}$ if compensation is accounted for, assuming no back-mutations). This yields a complete agent-based model of a stochastic exponential growth. 

Finally, in some simulations (see Fig.~\ref{FigMixedClust}), we explicitly took into account the possibility that mixed clusters form within immune hosts. For this, we employed our usual stochastic within-host growth scheme for the first $G-g$ generations of incubation, and we then switched to an explicit model of cluster formation for the last $g$ generations, where the cluster size (assumed to be equal to the bottleneck size) satisfies $N=2^g$. The last $g$ generations of incubation correspond to the formation of the transmitted clusters. We then formed clusters from each single bacterium present after $G-g$ generations of incubation. For these last $g$ generations, we assumed that bacteria have a fixed division time (and not a fixed division rate). As usual, at each division, daughter cells can mutate randomly with a given probability. We implemented this explicit construction of clusters only in the case where resistance has no cost ($\delta=0$), which is simpler because divisions within the fixed division time model are then synchronous.

\section{Analytical approximations for the spread probability without pre-existing resistance}\label{analyticalapprox}

\subsection{Proportion of resistant bacteria}\label{proportionmutants}

When starting with only sensitive bacteria, at the first round of replication, an average proportion $\mu_1 $ of resistant bacteria is produced. At the second round, neglecting back-mutations, and in the limit $\mu_1 \ll 1$, so that the proportion of mutant bacteria remains negligible, the proportion of mutant bacteria will be $\mu_1 (1-\delta)$ (the resistant bacteria produced at the first round, which reproduce more slowly than the sensitive ones), plus $\mu_1$ (the resistant bacteria newly produced). A similar reasoning can be applied for subsequent rounds of replication. Thus, after $G$ generations, the average proportion of resistant bacteria is
\begin{equation}
\mu_1 \left( (1-\delta)^{G-1} + (1-\delta)^{G-2} + ... + (1-\delta) + 1 \right) = \frac{\mu_1}{\delta} (1-(1-\delta)^G) . 
\label{propmut}
\end{equation}
Note that Eq.~\eqref{propmut} was obtained within a discrete description. Our continuous time model similarly yields $ (1-\exp(-G\delta))\mu_1/\delta$. In particular, if $G\delta\ll 1$, this yields a mutant proportion of $\mu_1 G$, while if $G\delta\gg 1$, this yields the mutation-selection balance proportion $\mu_1/\delta$.

\subsection{Approximation for a small number of generations}\label{smallGapproxsection}

\textbf{Regime considered}

As mentioned in the main text section titled ``Analytical approximations for the spread probability without pre-existing resistance'', here, we take into account mutations and the fitness cost of resistance, but not its compensation. We use the deterministic description of the within-host growth, and take $\lambda_\mathcal{N}=\lambda_\mathcal{I}=\lambda$, as well as  $q_\mathcal{N}=q_\mathcal{I}=q$. In addition, the present case of a small number $G$ of generations corresponds to $\delta G \ll 1$ and $(G-g) \ll 2 (N-1)$. Below, we explicitly demonstrate that if at least one resistant mutant is transmitted, the probability of extinction is similar for an immune and a naive host population. This is necessary to calculate the impact of clustering on spread probabilities in this regime (Eq.~\eqref{eqR1}).

~

$\mathbf{e_{N}^{\mathcal{N}}}$\textbf{ vs. }  $\mathbf{e_{N}^{\mathcal{I}}}$

Let us first compare the extinction probabilities starting from a host infected with only resistant bacteria in a fully naive host population $e_{N}^{\mathcal{N}}$ and in a fully immune one $e_{N}^{\mathcal{I}}$. In the limit of a small number of generations $G$, a host initially infected with only resistant bacteria will have a very small proportion of sensitive bacteria at the transmission time even in the absence of treatment. Thus $e_{N}^{\mathcal{N}} \approx e_{N}^{\mathcal{I}} \approx e_N$. 

~

$\mathbf{e_{1}^{\mathcal{N}}}$\textbf{ vs. }  $\mathbf{e_{1}^{\mathcal{I}}}$

Next, let us compare the extinction probabilities starting from a host initially infected with one resistant bacteria and $N-1$ sensitive ones, in a fully naive and in a fully immune population, i.e. $e_{1}^{\mathcal{N}}$ and $e_{1}^{\mathcal{I}}$. Because $\delta G \ll 1$, the proportion of resistant bacteria at transmission is close to $1/N$. Let us assume that $N\gg 1$. 

$\mathbf{e_{1}^{\mathcal{N}}}$

Let us first consider a fully naive population. For a non-treated host, the probability that no resistant bacteria is present among $N$ bacteria transmitted is $(1-1/N)^N=\exp(N\log(1-1/N))\approx \exp(-1)$. The probability that one resistant bacteria among $N$ is transmitted is $N (1/N) (1-1/N)^{N-1} = \exp((N-1)\log(1-1/N)) \approx \exp(-1) \exp(1/N)  \approx \exp(-1) $. The probability to transmit 2 resistant bacteria among $N$ is $N(N-1)/2 (1/N)^2 (1-1/N)^{N-2} =(1-1/N)/2  \exp((N-2)\log(1-1/N)) \approx \exp(-1) \exp(2/N) (1-1/N)/2 \approx \exp(-1)/2 $. Thus the probability that more than 2 resistant bacteria are transmitted is smaller than 10\%: most cases in which more than 1 resistant bacteria are transmitted are cases in which few, and mainly 2 resistant bacteria are transmitted. Since the outcome is not very different whether 1 or 2 resistant bacteria are transmitted, let us assume that $e_{1}^{\mathcal{N}} \approx e_{2}^{\mathcal{N}}$ and let us neglect transmission of more resistant bacteria. Under these assumptions, when a host is initially infected with 1 resistant bacteria and $N-1$ sensitive ones, then with probability $q$, this host is treated, leading to a fully resistant infection, and with probability $1-q$, it is not treated, in which case for each of the transmissions (whose number is Poisson distributed with average $\lambda$), it has a probability close to $\exp(-1)$ to transmit only sensitive bacteria, which leads to extinction with a probability close to 1 (more precisely, 1 minus a small spread probability arising from mutations of the order of $\mu_1$, see section~\ref{proportionmutants}), and a probability close to $(1-\exp(-1))$ to transmit at least one resistant bacteria, which leads to extinction with a probability close to $ e_{1}^{\mathcal{N}}$. This results in:  
\begin{equation}
    e_{1}^{\mathcal{N}} \approx q e_N + (1-q) \exp(-\lambda (1-\exp(-1))(1- e_{1}^{\mathcal{N}}) )
    \label{e1Nappx}
\end{equation}

$\mathbf{e_{1}^{\mathcal{I}}}$

Let us now turn to a fully immune population, and assume that clusters comprise a single type of bacteria since here the inoculum is mixed. When an immune host is initially infected with 1 resistant bacteria and $N-1$ sensitive bacteria, then with probability $q$, this host is treated, leading to a fully resistant infection, and with probability $1-q$, it is not treated, in which case for each of the transmissions (whose number is Poisson distributed with average $\lambda$), it has a probability $1-1/N$ to transmit only sensitive bacteria, which leads to an extinction probability close to 1 (more precisely, 1 minus a small spread probability arising from mutations of the order of $\mu_1$, see section~\ref{proportionmutants}), and a probability $1/N$ to transmit resistant bacteria only, leading to extinction with probability $ e_N$. This results in:   
\begin{equation}
    e_{1}^{\mathcal{I}} \approx q e_N + (1-q) \exp(-\lambda (1- e_N)/N )
    \label{e1Iappx}
\end{equation}

\textbf{Conclusion on} $\mathbf{e_{1}^{\mathcal{N}}}$\textbf{ vs. }  $\mathbf{e_{1}^{\mathcal{I}}}$

Since we assumed $N\gg 1$, $\exp(-\lambda (1- e_N)/N )$ will be of the order of 1, so Eqs.~\eqref{e1Nappx} and~\eqref{e1Iappx}  entail  $ e_{1}^{\mathcal{N}} < e_{1}^\mathcal{I}$.  Note that numerically, they remain of the same order of magnitude. For instance, in the limit of large $N$, the ratio of extinction probabilities $e_{1}^\mathcal{I}/ e_{1}^{\mathcal{N}}$ is at most 2 for $\lambda=2$. For instance in Fig.~\ref{Fig1}, this ratio is close to 2.

~

\textbf{Conclusion on }$\mathbf{e_{0}^{\mathcal{N}}}$\textbf{ vs. }  $\mathbf{e_{0}^{\mathcal{I}}}$

As explained in the main text, in a fully naive population, starting from a host infected with only sensitive bacteria, the probability for this host to transmit at least one resistant bacteria is $\approx\mu_1 G N$ (the proportion of resistant bacteria multiplied by the bottleneck size). Then the associated spread probability is $1-e_{1}^{\mathcal{N}}$. In contrast, in a fully immune population, starting from a host infected with only sensitive bacteria, the probability to transmit at least one resistant bacteria is approximately $2 \mu_1 (N-1)$, the proportion of mixed clusters. Half of these mixed clusters bear only one resistant bacteria. A quarter of them comprise 2 resistant bacteria, and so on, so we can approximate that the corresponding spread probability by  $1-e_{1}^{\mathcal{I}}$. 

Combining our results yields:
\begin{equation}\label{eqGratio}
 \mathcal{R}=\frac{1-e_0^\mathcal{N}}{1-e_0^\mathcal{I}}\approx \frac{G}{2}\frac{1-e_1^\mathcal{N}}{1-e_1^\mathcal{I}} \gtrsim\frac{G}{2}.
\end{equation}
This quantifies the impact of clustering on the spread of resistance in the case of a short incubation time. Note that in this regime, $G$ is not very large, but it can still be much larger than 2, in which case Eq.~\eqref{eqGratio} shows that the impact of clustering can be large.


\newpage

\begin{thebibliography}{54}
\providecommand{\natexlab}[1]{#1}
\providecommand{\url}[1]{\texttt{#1}}
\providecommand{\urlprefix}{URL }

\bibitem[{Andersson and Hughes(2010)}]{Andersson10}
Andersson, D.~I. and D.~Hughes, 2010.
\newblock {Antibiotic resistance and its cost: is it possible to reverse
  resistance?}
\newblock Nat. Rev. Microbiol. 8:260--271.

\bibitem[{Andr{\'{e}} and Day(2005)}]{Andre2005}
Andr{\'{e}}, J.-B. and T.~Day, 2005.
\newblock {The Effect of Disease Life History on the Evolutionary Emergence of
  Novel Pathogens}.
\newblock Proceedings of the Royal Society B: Biological Sciences
  272:1949--1956.

\bibitem[{Antia et~al.(2003)Antia, Regoes, Koella, and Bergstrom}]{Antia2003}
Antia, R., R.~R. Regoes, J.~C. Koella, and C.~T. Bergstrom, 2003.
\newblock {The role of evolution in the emergence of infectious diseases}.
\newblock Nature 426:658--661.

\bibitem[{Bansept et~al.(2018)Bansept, Moor-Schumann, Diard, Hardt, Slack, and
  Loverdo}]{bansept2018enchained}
Bansept, F., K.~Moor-Schumann, M.~Diard, W.-D. Hardt, E.~Slack, and C.~Loverdo,
  2018.
\newblock Enchained growth and cluster dislocation: a possible mechanism for
  microbiota homeostasis.
\newblock bioRxiv P. 298059.

\bibitem[{Carlet and Le~Coz(2015)}]{carlet2015rapport}
Carlet, J. and P.~Le~Coz, 2015.
\newblock Rapport du groupe de travail sp{\'e}cial pour la pr{\'e}servation des
  antibiotiques.
\newblock Tech. rep., Minist{\`e}re des Affaires sociales, de la Sant{\'e} et
  des Droits des femmes.
\newblock
  \urlprefix\url{http://solidarites-sante.gouv.fr/IMG/pdf/rapport\_antibiotiques.pdf}.

\bibitem[{Dahms et~al.(2014)Dahms, H{\"u}bner, Wilke, and
  Kramer}]{dahms2014mini}
Dahms, C., N.-O. H{\"u}bner, F.~Wilke, and A.~Kramer, 2014.
\newblock Mini-review: Epidemiology and zoonotic potential of multiresistant
  bacteria and clostridium difficile in livestock and food.
\newblock GMS hygiene and infection control 9.

\bibitem[{Desin et~al.(2013)Desin, K{\"o}ster, and
  Potter}]{desin2013salmonella}
Desin, T.~S., W.~K{\"o}ster, and A.~A. Potter, 2013.
\newblock Salmonella vaccines in poultry: past, present and future.
\newblock Expert review of vaccines 12:87--96.

\bibitem[{Diard et~al.(2017)Diard, Bakkeren, Cornuault, Moor, Hausmann, Sellin,
  Loverdo, Aertsen, Ackermann, De~Paepe, Slack, and
  Hardt}]{diard2017inflammation}
Diard, M., E.~Bakkeren, J.~K. Cornuault, K.~Moor, A.~Hausmann, M.~E. Sellin,
  C.~Loverdo, A.~Aertsen, M.~Ackermann, M.~De~Paepe, E.~Slack, and W.-D. Hardt,
  2017.
\newblock Inflammation boosts bacteriophage transfer between salmonella spp.
\newblock Science 355:1211--1215.

\bibitem[{Donaldson et~al.(2016)Donaldson, Lee, and
  Mazmanian}]{donaldson2016gut}
Donaldson, G.~P., S.~M. Lee, and S.~K. Mazmanian, 2016.
\newblock Gut biogeography of the bacterial microbiota.
\newblock Nature Reviews Microbiology 14:20.

\bibitem[{Evans and Calderwood(2007)}]{evans2007forces}
Evans, E.~A. and D.~A. Calderwood, 2007.
\newblock Forces and bond dynamics in cell adhesion.
\newblock Science 316:1148--1153.

\bibitem[{Ewens(1979)}]{Ewens79}
Ewens, W.~J., 1979.
\newblock {Mathematical Population Genetics}.
\newblock Springer-Verlag.

\bibitem[{Ewens(2012)}]{ewens2012mathematical}
---{}---{}---, 2012.
\newblock Mathematical population genetics 1: theoretical introduction,
  vol.~27.
\newblock Springer Science \& Business Media.

\bibitem[{Fournier and Parkos(2012)}]{fournier2012role}
Fournier, B. and C.~Parkos, 2012.
\newblock The role of neutrophils during intestinal inflammation.
\newblock Mucosal immunology 5:354.

\bibitem[{Frank and Slatkin(1990)}]{frank1990evolution}
Frank, S.~A. and M.~Slatkin, 1990.
\newblock Evolution in a variable environment.
\newblock The American Naturalist 136:244--260.

\bibitem[{Gillespie(1976)}]{Gillespie76}
Gillespie, D.~T., 1976.
\newblock {A general method for numerically simulating the stochastic time
  evolution of coupled chemical reactions}.
\newblock J. Comput. Phys. 22:403--434.

\bibitem[{Gillespie(1974)}]{gillespie1974natural}
Gillespie, J.~H., 1974.
\newblock Natural selection for within-generation variance in offspring number.
\newblock Genetics 76:601--606.

\bibitem[{Harris(1963)}]{Harris1963}
Harris, T.~E., 1963.
\newblock {The Theory of Branching Processes}.

\bibitem[{Harrison et~al.(2017)Harrison, Coll, Toleman, Blane, Brown,
  T{\"o}r{\"o}k, Parkhill, and Peacock}]{harrison2017genomic}
Harrison, E.~M., F.~Coll, M.~S. Toleman, B.~Blane, N.~M. Brown, M.~E.
  T{\"o}r{\"o}k, J.~Parkhill, and S.~J. Peacock, 2017.
\newblock Genomic surveillance reveals low prevalence of livestock-associated
  methicillin-resistant staphylococcus aureus in the east of england.
\newblock Scientific reports 7:7406.

\bibitem[{Hughes and Andersson(2015)}]{Hughes15}
Hughes, D. and D.~I. Andersson, 2015.
\newblock {{E}volutionary consequences of drug resistance: shared principles
  across diverse targets and organisms}.
\newblock Nat. Rev. Genet. 16:459--471.

\bibitem[{Iwasa et~al.(2004)Iwasa, Michor, and Nowak}]{Iwasa2004}
Iwasa, Y., F.~Michor, and M.~A. Nowak, 2004.
\newblock {Evolutionary dynamics of invasion and escape}.
\newblock Journal of Theoretical Biology 226:205--214.

\bibitem[{Jansen et~al.(2018)Jansen, Knirsch, and Anderson}]{Jansen18}
Jansen, K.~U., C.~Knirsch, and A.~S. Anderson, 2018.
\newblock {{T}he role of vaccines in preventing bacterial antimicrobial
  resistance}.
\newblock Nat. Med. 24:10--19.

\bibitem[{chaired~by Jim~O’Neill(2016)}]{AMR}
chaired~by Jim~O’Neill, ., 2016.
\newblock UK Review on Antimicrobial Resistance.

\bibitem[{Kendall(1948)}]{kendall1948generalized}
Kendall, D.~G., 1948.
\newblock On the generalized "birth-and-death" process.
\newblock The annals of mathematical statistics 19:1--15.

\bibitem[{Landers et~al.(2012)Landers, Cohen, Wittum, and
  Larson}]{landers2012review}
Landers, T.~F., B.~Cohen, T.~E. Wittum, and E.~L. Larson, 2012.
\newblock A review of antibiotic use in food animals: perspective, policy, and
  potential.
\newblock Public health reports 127:4--22.

\bibitem[{Lange(2010)}]{lange2010applied}
Lange, K., 2010.
\newblock Applied probability.
\newblock Springer Science \& Business Media.

\bibitem[{LeClair and Wahl(2018)}]{leclair2018impact}
LeClair, J.~S. and L.~M. Wahl, 2018.
\newblock The impact of population bottlenecks on microbial adaptation.
\newblock Journal of Statistical Physics 172:114--125.

\bibitem[{Levin et~al.(2000)Levin, Perrot, and Walker}]{Levin00}
Levin, B.~R., V.~Perrot, and N.~Walker, 2000.
\newblock {{C}ompensatory mutations, antibiotic resistance and the population
  genetics of adaptive evolution in bacteria}.
\newblock Genetics 154:985--997.

\bibitem[{Lipsitch and Siber(2016)}]{Lipsitch16}
Lipsitch, M. and G.~R. Siber, 2016.
\newblock {{H}ow {C}an {V}accines {C}ontribute to {S}olving the {A}ntimicrobial
  {R}esistance {P}roblem?}
\newblock MBio 7.

\bibitem[{Loverdo et~al.(2012)Loverdo, Park, Schreiber, and
  Lloyd-Smith}]{virus_life}
Loverdo, C., M.~Park, S.~Schreiber, and J.~O. Lloyd-Smith, 2012.
\newblock Influence of viral replication mechanisms on within-host evolutionary
  dynamics.
\newblock Evolution 66:3462--3471.

\bibitem[{Lynch(2010)}]{lynch2010evolution}
Lynch, M., 2010.
\newblock Evolution of the mutation rate.
\newblock TRENDS in Genetics 26:345--352.

\bibitem[{Maisnier-Patin and Andersson(2004)}]{maisnier2004adaptation}
Maisnier-Patin, S. and D.~I. Andersson, 2004.
\newblock Adaptation to the deleterious effects of antimicrobial drug
  resistance mutations by compensatory evolution.
\newblock Research in microbiology 155:360--369.

\bibitem[{Marrec and Bitbol(2018)}]{Marrec18}
Marrec, L. and A.~F. Bitbol, 2018.
\newblock {{Q}uantifying the impact of a periodic presence of antimicrobial on
  resistance evolution in a homogeneous microbial population of fixed size}.
\newblock J. Theor. Biol. 457:190--198.

\bibitem[{Mason(1935)}]{mason1935comparison}
Mason, M.~M., 1935.
\newblock A comparison of the maximal growth rates of various bacteria under
  optimal conditions.
\newblock Journal of bacteriology 29:103.

\bibitem[{McClure and Day(2014)}]{mcclure2014theoretical}
McClure, N.~S. and T.~Day, 2014.
\newblock A theoretical examination of the relative importance of evolution
  management and drug development for managing resistance.
\newblock Proceedings of the Royal Society of London B: Biological Sciences
  281:20141861.

\bibitem[{McGrady and Ziff(1987)}]{mcgrady1987shattering}
McGrady, E.~D. and R.~M. Ziff, 1987.
\newblock ‘‘shattering’’transition in fragmentation.
\newblock Physical review letters 58:892.

\bibitem[{Moor et~al.(2017)Moor, Diard, Sellin, Felmy, Wotzka, Toska, Bakkeren,
  Arnoldini, Bansept, Dal~Co, V{\"{o}}ller, Minola, Fernandez-Rodriguez,
  Agatic, Barbieri, Piccoli, Casiraghi, Corti, Lanzavecchia, Regoes, Loverdo,
  Stocker, Brumley, Hardt, and Slack}]{Moor2017}
Moor, K., M.~Diard, M.~E. Sellin, B.~Felmy, S.~Y. Wotzka, A.~Toska,
  E.~Bakkeren, M.~Arnoldini, F.~Bansept, A.~Dal~Co, T.~V{\"{o}}ller, A.~Minola,
  B.~Fernandez-Rodriguez, G.~Agatic, S.~Barbieri, L.~Piccoli, C.~Casiraghi,
  D.~Corti, A.~Lanzavecchia, R.~R. Regoes, C.~Loverdo, R.~Stocker, D.~R.
  Brumley, W.-D. Hardt, and E.~Slack, 2017.
\newblock {High-avidity IgA protects the intestine by enchaining growing
  bacteria}.
\newblock Nature 544:498--502.

\bibitem[{Odell and Keller(1986)}]{odell1986flow}
Odell, J.~A. and A.~Keller, 1986.
\newblock Flow-induced chain fracture of isolated linear macromolecules in
  solution.
\newblock Journal of Polymer Science Part B: Polymer Physics 24:1889--1916.

\bibitem[{Orr and Unckless(2008)}]{Orr2008}
Orr, H.~A. and R.~L. Unckless, 2008.
\newblock {Population extinction and the genetics of adaptation}.
\newblock The American Naturalist 172:160--9.

\bibitem[{Park et~al.(2013)Park, Loverdo, Schreiber, and
  Lloyd-Smith}]{miran_cross}
Park, M., C.~Loverdo, S.~J. Schreiber, and J.~O. Lloyd-Smith, 2013.
\newblock Multiple scales of selection influence the evolutionary emergence of
  novel pathogens.
\newblock Philosophical Transactions B P. 20120333.

\bibitem[{Paulander et~al.(2007)Paulander, Maisnier-Patin, and
  Andersson}]{Paulander07}
Paulander, W., S.~Maisnier-Patin, and D.~I. Andersson, 2007.
\newblock {{M}ultiple mechanisms to ameliorate the fitness burden of mupirocin
  resistance in {S}almonella typhimurium}.
\newblock Mol. Microbiol. 64:1038--1048.

\bibitem[{Poon et~al.(2005)Poon, Davis, and Chao}]{Poon05}
Poon, A., B.~H. Davis, and L.~Chao, 2005.
\newblock {{T}he coupon collector and the suppressor mutation: estimating the
  number of compensatory mutations by maximum likelihood}.
\newblock Genetics 170:1323--1332.

\bibitem[{Rhomberg and Jones(2009)}]{rhomberg2009summary}
Rhomberg, P.~R. and R.~N. Jones, 2009.
\newblock Summary trends for the meropenem yearly susceptibility test
  information collection program: a 10-year experience in the united states
  (1999--2008).
\newblock Diagnostic microbiology and infectious disease 65:414--426.

\bibitem[{Schrag et~al.(1997)Schrag, Perrot, and Levin}]{Schrag97}
Schrag, S.~J., V.~Perrot, and B.~R. Levin, 1997.
\newblock {Adaptation to the fitness cost of antibiotic resistance in
  \textit{E. coli}}.
\newblock Proc. R. Soc. Lond. B 264:1287--1291.

\bibitem[{Schreiber et~al.(2016)Schreiber, Ke, Loverdo, Park, Ahsan, and
  Lloyd-Smith}]{schreiber2016crossscale}
Schreiber, S.~J., R.~Ke, C.~Loverdo, M.~Park, P.~Ahsan, and J.~O. Lloyd-Smith,
  2016.
\newblock Crossscale dynamics and the evolutionary emergence of infectious
  diseases. biorxiv .

\bibitem[{Sender et~al.(2016)Sender, Fuchs, and Milo}]{sender2016revised}
Sender, R., S.~Fuchs, and R.~Milo, 2016.
\newblock Revised estimates for the number of human and bacteria cells in the
  body.
\newblock PLoS biology 14:e1002533.

\bibitem[{Sharma et~al.(2018)Sharma, Caraguel, Sexton, McWhorter, Underwood,
  Holden, and Chousalkar}]{sharma2018shedding}
Sharma, P., C.~Caraguel, M.~Sexton, A.~McWhorter, G.~Underwood, K.~Holden, and
  K.~Chousalkar, 2018.
\newblock Shedding of salmonella typhimurium in vaccinated and unvaccinated
  hens during early lay in field conditions: a randomised controlled trial.
\newblock BMC microbiology 18:78.

\bibitem[{Moura~de Sousa et~al.(2015)Moura~de Sousa, Sousa, Bourgard, and
  Gordo}]{deSousa15}
Moura~de Sousa, J., A.~Sousa, C.~Bourgard, and I.~Gordo, 2015.
\newblock {{P}otential for adaptation overrides cost of resistance}.
\newblock Future Microbiol 10:1415--1431.

\bibitem[{Stecher and Hardt(2008)}]{stecher2008role}
Stecher, B. and W.-D. Hardt, 2008.
\newblock The role of microbiota in infectious disease.
\newblock Trends in microbiology 16:107--114.

\bibitem[{{Sécurité Sociale}(2014)}]{consultations}
{Sécurité Sociale}, 2014.
\newblock Projet de loi de financement de la sécurité sociale - annexe 1 :
  Programmes de qualité et d’efficience - programme de qualité et
  d’efficience « maladie » - sous-indicateur numéro 9-2.
\newblock Tech. rep., Sécurité sociale.
\newblock
  \urlprefix\url{http://www.securite-sociale.fr/IMG/pdf/plfss14\_annexe1\_pqe\_maladie\_indicateur9.pdf}.

\bibitem[{Tees et~al.(1993)Tees, Coenen, and Goldsmith}]{Tees1993}
Tees, D.~F., O.~Coenen, and H.~L. Goldsmith, 1993.
\newblock {Interaction forces between red cells agglutinated by antibody. IV.
  Time and force dependence of break-up}.
\newblock Biophysical Journal 65:1318--1334.

\bibitem[{Traulsen and Hauert({2009})}]{Traulsen09}
Traulsen, A. and C.~Hauert, {2009}.
\newblock {Stochastic evolutionary game dynamics}.
\newblock \emph{in} H.-G. Schuster, ed. {Reviews of Nonlinear Dynamics and
  Complexity}, vol.~{II}. Wiley-VCH.

\bibitem[{Wienand et~al.(2015)Wienand, Lechner, Becker, Jung, and
  Frey}]{Wienand15}
Wienand, K., M.~Lechner, F.~Becker, H.~Jung, and E.~Frey, 2015.
\newblock {{N}on-{S}elective {E}volution of {G}rowing {P}opulations}.
\newblock PLoS ONE 10:e0134300.

\bibitem[{zur Wiesch et~al.(2011)zur Wiesch, Kouyos, Engelstadter, Regoes, and
  Bonhoeffer}]{zurWiesch11}
zur Wiesch, P.~A., R.~Kouyos, J.~Engelstadter, R.~R. Regoes, and S.~Bonhoeffer,
  2011.
\newblock {{P}opulation biological principles of drug-resistance evolution in
  infectious diseases}.
\newblock Lancet Infect Dis 11:236--247.

\bibitem[{{World Health Organization \& Food and Agriculture
  Organization}(2002)}]{WHO105}
{World Health Organization \& Food and Agriculture Organization}, 2002.
\newblock Risk assessments of Salmonella in eggs and broiler chickens, vol.
  http://www.who.int/foodsafety/publications/micro/salmonella/en/.

\end{thebibliography}
\end{document}